\documentclass[twocolumn,showpacs,preprintnumbers,amsmath,amssymb,floatfix]{revtex4}

\usepackage{graphicx}
\usepackage{dcolumn}
\usepackage{bm}

\usepackage{amsmath,amssymb}

\newcommand{\re}{\text{Re}}
\newcommand{\im}{\text{Im}}

\newcommand{\mk}{\frac{m_K^2}{f^2}}
\newcommand{\mpi}{\frac{m_{\pi}^2}{f^2}}

\newcommand{\Kpi}{\frac{m_K^2+m_\pi^2}{2f^2}}
\newcommand{\Kcpi}{\frac{5m_K^2-3m_\pi^2}{2f^2}}
\newcommand{\Kopi}{\frac{8m_K^2-5m_\pi^2}{3f^2}}

\newcommand{\ra}{\rangle}
\newcommand{\la}{\langle}

\newcommand{\be}{\begin{equation}}
\newcommand{\ee}{\end{equation}}
\newcommand{\ba}{\begin{eqnarray}}
\newcommand{\ea}{\end{eqnarray}}

\begin{document}

\preprint{}

\title{Two-meson cloud contribution to the baryon
antidecuplet binding}

\author{A. Hosaka$^{1}$, T. Hyodo$^{1 *}$\footnotetext{$^*$Electronic address: 
hyodo@rcnp.osaka-u.ac.jp}, 
F.J.~Llanes-Estrada$^{2 \dagger}$\footnotetext{$^\dagger$Electronic address: 
fllanes@fis.ucm.es},
E.~Oset$^3$, J.~R.~Pel\'aez$^4$, and M.~J.~Vicente~Vacas$^3$}

\affiliation{$^1$Research Center for Nuclear Physics (RCNP),
Ibaraki, Osaka 567-0047, Japan.\\
$^2$Universidad Complutense de Madrid, Depto. F\'{\i}sica
Te\'orica I, 28040 Madrid, Spain.\\
$^3$Departamento de F\'isica Te\'orica and IFIC,
Centro Mixto Universidad de Valencia-CSIC,
Institutos de Investigaci\'on de Paterna, Aptd. 22085, 46071
Valencia, Spain. \\
$^4$Universidad Complutense de Madrid, Depto. F\'{\i}sica
Te\'orica II, 28040 Madrid, Spain.
}

\date{\today}

\begin{abstract}
    We study the two-meson virtual cloud contribution to the
    self-energy of the SU(3) antidecuplet, to which the $\Theta^+$
    pentaquark is assumed to belong.
    This is motivated by the large branching ratio of the $N(1710)$
    decay into two pions and one nucleon.
    We derive effective Lagrangians
    that describe the $N(1710)$ decay into $N\pi\pi$ with
    two pions in $s$ or $p$ wave.
    We obtain increased binding 
    for all members of the antidecuplet and a contribution to the mass
    splitting between states with different strangeness which is at least 
    20\% of the empirical one.
    We also provide predictions for three-body decays of the pentaquark
    antidecuplet.
\end{abstract}

\maketitle

\section{Introduction}

The observation of the signal of the $\Theta^+$ exotic
baryon~\cite{Nakano:2003qx}, broadly known
as the pentaquark for its minimal five-quark Fock space assignment in
quantum chromodynamics, was stimulated
by the prediction~\cite{Diakonov:1997mm,Prasza}
of a $1/2^+$ baryon antidecuplet.
Many experimental and theoretical studies have been devoted to this
resonance~\cite{hyodoweb,Penta04}.
The original states proposed to form this
antidecuplet~\cite{Diakonov:1997mm}
were:
\begin{equation}
    \Theta(1540), \ N(1710), \ \Sigma(1890), \ \Xi(2070) ,
    \nonumber
\end{equation}
where the first and last, explicitly exotic states,
had not been observed at that time.
Isospin SU(2) is expected to hold to very good accuracy
and we do not list the $I_3$ quantum number.

This assignment is now, however,  challenged for
several reasons.
First, the NA49 collaboration~\cite{Alt:2003vb} reported
evidence for an exotic cascade $\Xi^{--}$,
probably in the same antidecuplet,
with a much lighter mass, 1860 MeV.
This is somewhat problematic
as doubts have arisen~\cite{Fischer:2004qb},
and it has not been seen in other
experiments~\cite{Knopfle:2004tu,:2004yk}
(See, however, K. Kadija's presentation at the
PENTAQUARK04 Workshop~\cite{Kadija}
with new reanalyses still supporting the
findings of Ref.~\cite{Fischer:2004qb}).
However, should the state be reconfirmed at 1860 MeV,
using the standard Gell-Mann--Okubo rule (GMO) of equal mass
splittings for the SU(3) antidecuplet,
the mass of $N_{\overline{10}}$ would have to be near
1647 MeV, which is about 60 MeV below the nominal one
$N(1710)$.
Furthermore, the mass of $\Sigma_{\overline{10}}$ would
have to be about $1753$ MeV.
Since a $\Sigma$ resonance is listed at $1770$ MeV with the
same spin and parity,
we will refer to the $\Sigma$ member of the antidecuplet
as $\Sigma(1770)$.
The association of this state to the $\Sigma(1890)$ 
would not fit in that scenario~\cite{Diakonov:2003jj}.

Second, quark model calculations that have appeared
after the report of the evidence of $\Theta^+$ tend
to predict an $N_{\overline{10}}$ at 
around $1650$ MeV~\cite{Carlson:2003pn,Carlson:2003wc}.
These predicted the
$\Xi_{\overline{10}}$ at $1900$ MeV~\cite{Carlson:2003wc},
which is
more in line with the experimental
outcome than the original calculation of the chiral soliton model,
although the latter can be readjusted (then underestimating the $N^*$).

Another difficulty arises because of the potential mixing of the nonexotic
members of the multiplet, the $N_{\overline{10}}$ and
$\Sigma_{\overline{10}}$
with members of pentaquark or ordinary three-quark
octets~\cite{Diakonov:2003jj,Jaffe:2003sg,Oh:2003fs}.
This would make the mass splitting between the physical states
dependent on two mixing angles.  A current conjecture is a mixing with the
Roper resonance~\cite{Jaffe:2003sg} that would, by level repulsion,
push the $N(1710)$ further above the $\Theta^+$ than predicted by the GMO
rule. Also with ideal mixing, the hidden strangeness $s\bar{s}$
wave function dominates the $N(1710)$, thus raising its mass.
However, such a strong mixing is not preferred by other
authors~\cite{Diakonov:2003jj,Cohen:2004gu,Pakvasa:2004pg}.
In summary, a new $N^*$ state would have to be searched 
for at a smaller mass if we were to impose perfect GMO rule.

The models we work with in this paper are rather phenomenological.
However, our method, based on symmetry principles,
is suited to at least estimating meson cloud effects,
which are important for the understanding of pentaquark properties. 
The main conclusion of this work is that the
virtual ``two-meson cloud'' yields an attractive
self-energy that provides about 20\% of the pentaquark mass splittings.
We believe that our study here
will become useful when more data are available.

The study presented here is complementary and looks for another 
source of mass splitting not contemplated by the GMO rule.
It would come from the two-meson
cloud. The possibility of constructing the $\Theta^+$ as a $K\pi N$ bound
state~\cite{Bicudo:2003rw,Kishimoto:2003xy,Bicudo:2004cm,Huang:2004}
has been examined in some detail~\cite{Llanes-Estrada:2003us}
employing meson-meson and meson-baryon interactions from chiral
Lagrangians, where attraction was found but not strong enough to bind the
system. Yet, this result leaves one wondering as to what role the
two-meson cloud could play in the stability of the state.
Coupling to multi-meson components is also implicit in the chiral
soliton picture, which leads to small masses of the
$\Theta^+$~\cite{Diakonov:1997mm,Kochelev:2004nd}.

In the present paper, we do not face the possible contribution
of the one-meson cloud to the antidecuplet binding, which can be
easily addressed as a minor correction to our results.
The small width of the $\Theta^+$ to $KN$, in spite of the
appreciable phase space available, qualitatively demands that this
contribution should be reasonably small;
in fact, it has been checked quantitatively in
Refs.~\cite{Kim:2004fk,Mohta:2004yf,Cabrera:2004yg}.
The self-energy of $\Theta^+$ with a two-meson cloud has been 
studied in parallel~\cite{Cabrera:2004yg}
in the context of the medium modification of $\Theta^+$ and
possible formation of $\Theta^+$ hypernuclei~\cite{Nagahiro:2004wu}.
We here report in full on vacuum results for not only
the $\Theta^+$ but also other members of the antidecuplet.

An important experimental input relevant to the present
study is the relatively large branching ratio
of $N(1710)$ into $N\pi\pi$, about $40-90\%$~\cite{Eidelman:2004wy}.
The branching
ratio into $N\pi\pi$ with the two pions in an $s$ wave is $10-40\%$ and
into $\rho N$, $5-25\%$.
This $N(1710)$ resonance and its
baryon-meson-meson decay mode has
been used in Ref.~\cite{Hyodo:2003jw} 
to produce a good shape of the $\Sigma \pi$
distribution in the $\pi^- p\to K^0\pi\Sigma$
reaction leading to the $\Lambda(1405)$~\footnote{In that work,
the N(1710) was assumed to belong to an octet representation,
although such a selection is not crucial to the results obtained 
there.}.

In the present work, we assume
that the $N(1710)$ has a large
antidecuplet component~\cite{Diakonov:2003jj}, and we will
perform a study of the $N\pi\pi$ $s$ wave and $\rho N$ decay
channels of this resonance
and their influence on the masses of various members of the antidecuplet.
Certainly one has to accept a mixing
with an octet component for realistic resonances in order,
for instance,  to explain the $N(1710)$  decay into $\Delta \pi$,
which is
forbidden for its antidecuplet component~\cite{Oh:2003fs,Close:2004tp}.
But we do not expect the mixing angle
to be close to ideal, as this would imply a stronger
$\Lambda K$ branching ratio than
$5-25\%$, as observed experimentally.
The decay pattern
of $N(1710)$ and $N(1440)$ also supports
the small mixing angle~\cite{Cohen:2004gu,Pakvasa:2004pg}.

The present study also provides information on the 
antidecuplet baryon-baryon-meson-meson $(PBMM)$ contact
interaction, which could be applied to the study of $\Theta^+$
production with the $\pi^{-}p \to K^{-}\Theta^+$
and $K^+p \to \pi^{+}\Theta^+$ reactions.
These reactions are studied in Refs.~\cite{Hyodo:2003th,Liu:2003rh,
Oh:2003gj,Oh:2003kw} and experimental information is becoming
available~\cite{Imai,Miwa}.

The paper is organized as follows.
In Sec.~\ref{sec:Lag}, we construct various $PBMM$
interactions with the two octet mesons and one baryon belonging
to octets and with the other baryon to an antidecuplet.
In Sec.~\ref{sec:Self}, we compute the contributions
of two-meson and one-baryon
loops to the mass splittings among the members of antidecuplet baryons.
In Sec.~\ref{sec:Results}, we present numerical results 
and discuss the importance
of two-meson contributions to the mass splittings
and partial decay widths.
As we will see, the contributions from the two-meson loops
provide sizable contributions to supplement the mass splittings 
naively expected from strange quark counting.
We will then discuss the range of interaction strengths of various
coupling terms. Section~\ref{sec:Summary} is devoted to a summary.
We also add appendices, where complete tables for the
$PBMM$ interactions are presented.

\section{Construction of effective interaction Lagrangians}\label{sec:Lag}
\subsection{Definition of fields}\label{subsec:Def}

Following a common convention~\cite{Ecker:1995gg,Pich:1995bw,Bernard:1995dp},
we write the physical meson and baryon fields as follows
\begin{align}
    \phi=&
    \begin{pmatrix}
        \frac{1}{\sqrt{2}}\pi^{0}+\frac{1}{\sqrt{6}}\eta & 
        \pi^{+} & K^{+} \\ 
        \pi^{-} & -\frac{1}{\sqrt{2}}\pi^{0}
        +\frac{1}{\sqrt{6}}\eta & K^{0} \\ 
        K^{-} & \bar{K}^{0} & -\frac{2}{\sqrt{6}}\eta
    \end{pmatrix} ,
    \label{eq:mesonLag} \\
    B=&
    \begin{pmatrix}
        \frac{1}{\sqrt{2}}\Sigma^{0}+\frac{1}{\sqrt{6}}\Lambda & 
        \Sigma^{+} & p \\ 
        \Sigma^{-} & -\frac{1}{\sqrt{2}}\Sigma^{0}
        +\frac{1}{\sqrt{6}}\Lambda & n \\ 
        \Xi^{-} & \Xi^{0} & -\frac{2}{\sqrt{6}}\Lambda
    \end{pmatrix} . 
    \label{eq:baryonLag}
\end{align}
The antidecuplet containing the exotic pentaquark states is a tensor 
$P^{ijk}$ totally symmetric in its three SU(3) indices.
The components of $P^{ijk}$ are related to the physical fields by
\begin{align}
    P^{333} &= \sqrt{6}\Theta^{+}_{\overline{10}}  , &
    P^{133} &= \sqrt{2} \, N^0_{\overline{10}}  ,  \nonumber \\
    P^{233} &= -\sqrt{2} \, N^+_{\overline{10}}  ,&
    P^{113} &= \sqrt{2} \, \Sigma^{-}_{\overline{10}} ,\nonumber \\
    P^{123} &= -\Sigma^{0}_{\overline{10}} ,&
    P^{223} &= -\sqrt{2} \, \Sigma^{+}_{\overline{10}}  , 
    \label{eq:pentaLag} \\ 
    P^{111} &=  \sqrt{6}\Xi^{--}_{\overline{10}} , &
    P^{112} &= -\sqrt{2}  \, \Xi_{\overline{10}}^{-}  , \nonumber \\
    P^{122} &= \sqrt{2} \, \Xi_{\overline{10}}^{0} , & 
    P^{222} &= - \sqrt{6}\Xi_{\overline{10}}^{+}  ,\nonumber
\end{align}
where we have adopted the normalization in Ref.~\cite{Oh:2004gz},
which is different from those used in 
Refs.~\cite{Mohta:2004yf,Ko:2003xx,Liu:2004qx}
by a sign and/or a factor.

Now we consider the possible interaction Lagrangians, constrained to be 
SU(3) symmetric.
We intend to address the process 
\begin{equation}
    \bm{8}_M + \bm{8}_M + \bm{8}_B
    \to \overline{\bm{10}}_P  ,
    \label{eq:process}
\end{equation}
where an octet baryon $\bm{8}_{B}$ and two octet
mesons $\bm{8}_{M}$ couple to 
an antidecuplet baryon $\overline{\bm{10}}_P$.
To have an SU(3) invariant Lagrangian,
we couple first the two $\bm{8}_M$ and then combine the resulting
irreducible representations with the baryon $\bm{8}_B$
to produce a $\overline{\bm{10}}_{BMM}$ representation. 
The group theoretical irreducible decomposition gives
\begin{align}
    &\bm{8}_M\otimes\bm{8}_M\otimes\bm{8}_B \nonumber \\
    =&
    (\bm{1}\oplus\bm{8}^s\oplus\bm{8}^a\oplus\bm{10}
    \oplus\overline{\bm{10}}\oplus\bm{27}
    )_{MM}\otimes\bm{8}_B\nonumber \\
    =&
    \bm{8}    \quad \gets \text{from } \bm{1}_{MM}\otimes\bm{8}_B 
    \nonumber\\
    &\oplus 
    (\bm{1}\oplus\bm{8}\oplus\bm{8}\oplus\bm{10}
    \oplus\overline{\bm{10}}\oplus\bm{27})
    \quad \gets \text{from } \bm{8}^s_{MM}\otimes\bm{8}_B  \nonumber \\
    &\oplus 
    (\bm{1}\oplus\bm{8}\oplus\bm{8}\oplus\bm{10}
    \oplus\overline{\bm{10}}\oplus\bm{27}) 
    \quad \gets \text{from } \bm{8}^a_{MM}\otimes\bm{8}_B\nonumber \\
    &\oplus (
    \bm{8}\oplus\bm{10}
    \oplus\bm{27}\oplus\bm{35}
    ) \quad \gets \text{from } \bm{10}_{MM}\otimes\bm{8}_B\nonumber\\
    &\oplus (
    \bm{8}\oplus\overline{\bm{10}}
    \oplus\bm{27}\oplus\bm{35^{\prime}}
    ) \quad \gets \text{from } \overline{\bm{10}}_{MM}\otimes\bm{8}_B
    \nonumber\\  
    &\oplus (
    \bm{8}\oplus\bm{10}\oplus\overline{\bm{10}}
    \oplus\bm{27}\oplus\bm{27}\oplus\bm{35}\oplus\bm{35^{\prime\prime}}
    \oplus\bm{64})\nonumber \\
    &\quad \gets \text{from } \bm{27}_{MM}\otimes\bm{8}_B .
    \label{eq:decomposition}
\end{align}
Here $\bm{8}^s$ and $\bm{8}^a$ denote symmetric and antisymmetric
combinations of the two-meson fields.
Hence we obtain four $\overline{\bm{10}}_{BMM}$ representations after recoupling
$\bm{8}^s_{MM}$, $\bm{8}^a_{MM}$, $\bm{10}_{MM}$ and
$\bm{27}_{MM}$ with $\bm{8}_{B}$.

\subsection{Two-meson $\bm{8}^s$ representation}\label{subsec:8sLag}

In constructing effective Lagrangians, we follow the
principle of using the minimum numbers of derivatives in the
fields. This will be released later when we
discuss possible structures involving derivatives.
To construct $\bm{8}^s$ from
two $\bm{8}_M$, we have in tensor notation
\begin{align}
    D_i{}^{j}[\bm{8}^s_{MM}]
    =&\phi_{i}{}^{a}\phi_{a}{}^{j}+\phi_{i}{}^{a}\phi_{a}{}^{j} 
    -\frac{2}{3}\delta_{i}{}^{j}\phi_{a}{}^{b}\phi_{b}{}^{a}
    \nonumber \\
    =&2\phi_{i}{}^{a}\phi_{a}{}^{j} 
    -\frac{2}{3}\delta_{i}{}^{j}\phi_{a}{}^{b}\phi_{b}{}^{a} .
    \label{eq:MM8s}
\end{align}
We combine this now with an $\bm{8}_B$ to give an
antidecuplet
\begin{align}
    T^{ijk}[\overline{\bm{10}}_{BMM(8s)}]
    =&2\phi_{l}{}^{a}\phi_{a}{}^{i} B_{m}{}^{j}\epsilon^{lmk}
   \nonumber \\
    &+(i,j,k\text{ symmetrized})  .
    \label{eq:MM8sB}
\end{align}
Hence, the interaction Lagrangian becomes
\begin{equation}
    \mathcal{L}^{8s}
    =\frac{g^{8s}}{2f}
    \bar{P}_{ijk}\epsilon^{lmk}
    \phi_{l}{}^{a}\phi_{a}{}^{i} B_{m}{}^{j}+ \text{h.c.} ,
    \label{eq:8sLag}
\end{equation}
where h.c. denotes the Hermitian conjugate terms,
in order to take into account the processes
in which the antidecuplet is in the initial state.
Note also that two $\phi$ fields
have appeared, and we have included a factor
$1/2f$ in order to make
$g^{8s}$ dimensionless ($f$ is the pion decay constant
$f=93$ MeV).

\subsection{Two-meson $\bm{8}^a$ representation}\label{subsec:8aLag}

Next we take the antisymmetric combination of the $\bm{8}_M$
and $\bm{8}_M$, which for identical meson octets leads to
\begin{equation}
    A_i{}^{j}[\bm{8}^a_{MM}]
    =\phi_{i}{}^{a} \phi_{a}{}^{j}
    -\phi_{i}{}^{a} \phi_{a}{}^{j}=0 .
    \label{eq:MM8a}
\end{equation}
So given the identity of the meson octets,
this combination is zero.
The simplest way to construct the Lagrangian of this structure
is to introduce a derivative in one of the fields,
which leads automatically to the vector current consisting of 
two meson fields. Proceeding as before, 
we combine this structure with the $\bm{8}_B$
to give $\overline{\bm{10}}$ then finally
\begin{equation}
    \mathcal{L}^{8a}
    =i\frac{g^{8a}}{4f^2}
    \bar{P}_{ijk}\epsilon^{lmk}\gamma^{\mu}
    (\partial_{\mu}\phi_{l}{}^{a} \phi_{a}{}^{i}
    - \phi_{l}{}^{a}\partial_{\mu} \phi_{a}{}^{i})B_{m}{}^{j} + 
    \text{h.c.}
    ,
    \label{eq:8aLag}
\end{equation}
where 
$g^{8a}$ is dimensionless.
This interaction Lagrangian contains the coupling of the $N(1710)$
with $N\pi\pi$, the two pions in a $\rho$-meson type
correlation.
From the experimental branching ratio, we can determine the coupling
constant $g^{8a}$.

\subsection{Two-meson $\overline{\bm{10}}$ representation}\label{subsec:10Lag}

To construct the $\overline{\bm{10}}$ combination
from two mesons, we have now
\begin{align}
    T_{MM}^{ijk}[\overline{\bm{10}}_{MM}]
    =&  \epsilon^{lmk}\phi_l{}^{i}\phi_m{}^{j}
    +(i,j,k\text{ symmetrized}) \nonumber\\
    =& \epsilon^{lmk}\phi_l{}^{i}\phi_m{}^{j}
    +\epsilon^{lmk}\phi_l{}^{j}\phi_m{}^{i}
    +\epsilon^{lmi}\phi_l{}^{j}\phi_m{}^{k} \nonumber \\
    &
    +\epsilon^{lmi}\phi_l{}^{k}\phi_m{}^{j}
    +\epsilon^{lmj}\phi_l{}^{k}\phi_m{}^{i}
    +\epsilon^{lmj}\phi_l{}^{i}\phi_m{}^{k} \nonumber\\
    =&0 ,
    \label{eq:MMbar10}
\end{align}
which is identically zero for equal meson octets.

\subsection{Two-meson $\bm{27}$ representation}\label{subsec:27Lag}

The expansion for the $\bm{27}$ representation leads to
\begin{align}
    H_{ik}^{jl}[\bm{27}_{MM}]
    =
    &\phi_{i}{}^{j} \phi_{k}{}^{l}
    + \phi_{i}{}^{l} \phi_{k}{}^{j}
    +\phi_{k}{}^{j} \phi_{i}{}^{l}
    + \phi_{k}{}^{l} \phi_{i}{}^{j} \nonumber\\
    &-\frac{1}{5}
    \left(\delta_{i}{}^{j}D_{k}{}^{l}
    + \delta_{i}{}^{l} D_{k}{}^{j}
    +\delta_{k}{}^{j} D_{i}{}^{l}
    + \delta_{k}{}^{l} D_{i}{}^{j}\right)\nonumber \\
    &-\frac{1}{6}
    \left(\delta_{i}{}^{j}\delta_{k}{}^{l}
    \phi_{a}{}^{b} \phi_{b}{}^{a}
    + \delta_{i}{}^{l} \delta_{k}{}^{j}
    \phi_{a}{}^{b} \phi_{b}{}^{a}\right)  ,
    \label{eq:MM27}
\end{align}
where 
$D_{i}{}^{j}$ is defined in Eq.~\eqref{eq:MM8s}.
Now the combination of $\bm{27}_{MM}$ to $\bm{8}_B$
to give the $\overline{\bm{10}}$ representation leads to
\begin{align}
    \mathcal{L}^{27}
    =&\frac{g^{27}}{2f}
    \Bigl[4\bar{P}_{ijk}\epsilon^{lbk}
    \phi_{l}{}^{i} \phi_{a}{}^{j}B_{b}{}^{a} \nonumber\\
    &-\frac{4}{5}\bar{P}_{ijk}\epsilon^{lbk}
    \phi_{l}{}^{a} \phi_{a}{}^{j}
    B_{b}{}^{i}
    \Bigr] + \text{h.c.}  ,
    \label{eq:27Lag}
\end{align}
where the first term gives us a new SU(3) structure,
but the second one is equal to $\mathcal{L}^{8s}$
given in Eq.~\eqref{eq:8sLag}.

To summarize briefly, for the possible SU(3) symmetric
couplings of $PBMM$, there are two independent terms
with no derivatives, namely
Eqs.~\eqref{eq:8sLag} and \eqref{eq:27Lag}.
With  one derivative, there are four more
terms available, but we will consider only Eq.~\eqref{eq:8aLag},
which has the structure 
for the decay of $N(1710)\to N
\pi\pi(p$-wave) as observed experimentally.

\subsection{Chiral symmetric Lagrangians}\label{subsec:chiralLag}

In the perturbative chiral Lagrangian approach, one would like to 
implement chiral symmetry as a derivative expansion. 
In addition, one of the advantages of 
chiral Lagrangians is that they 
relate coupling constants 
of different processes and, in particular, with 
increasing number of mesons.  
However, in the present case we cannot take advantage of any of these 
relations, since the couplings for the present Lagrangians are 
\textit{a priori} 
completely arbitrary, and we are only interested in the two-meson 
problem.  
Still, in this section we build the lowest-order
chiral Lagrangian, with two derivatives.
Let us remark that the chiral expansion with baryons
is known to converge much more slowly than chiral perturbation
theory with mesons, and this lowest-order Lagrangian can 
only be expected
to give a mere qualitative description of the physics.
For that reason,
to build the Lagrangians of 
the previous Secs.~\ref{subsec:8sLag} 
and \ref{subsec:27Lag} we just relied on flavor SU(3).  
Still, we will check here that the lack of chiral symmetry in those
Lagrangians
does not have much relevance to the
mass splittings and decays we are interested in, since
already with the leading-order
Lagrangian we get qualitatively the same results.
In other words, the relevant symmetry here is SU(3), not chiral symmetry.

To show this, we write a chiral invariant
Lagrangian by making the substitution 
$\phi \cdot \phi \to A_\mu \cdot A^\mu$ in
Eq.~\eqref{eq:8sLag} such that 
\begin{equation}
    \mathcal{L}^{\chi}
    =\frac{g^{\chi}}{2f}
    \bar{P}_{ijk}\epsilon^{lmk}
    (A_{\mu})_{l}{}^{a}(A^{\mu})_{a}{}^{i} B_{m}{}^{j}+
    \text{h.c.}  ,
    \label{eq:chiralLag}
\end{equation}
where $A_\mu$ is the axial current written in terms of the 
chiral field $\xi$:
\begin{equation}
    A_{\mu} = \frac{i}{2}
      \left( \xi^\dagger \partial_\mu \xi -
        \xi \partial_\mu \xi^\dagger \right)
        ,   \label{eq:axial}
\end{equation}
with $\xi=e^{i\phi/\sqrt{2}f}$.
To the leading order in meson field, $A_\mu \sim -\partial_\mu \phi
/\sqrt{2}f$, we find the 
interaction Lagrangian
induced from Eq.~\eqref{eq:chiralLag} by making the replacement
\begin{equation}
    (A_{\mu})_{l}{}^{a}(A^{\mu})_{a}{}^{i} 
    \to \frac{1}{2f^2}\partial_\mu 
    \phi_{l}{}^{a}\partial^\mu \phi_{a}{}^{i} .
\end{equation}
Obviously, the SU(3) structure is not affected by this procedure,
although the use of Lagrangians involving derivatives will
introduce some degree of SU(3) breaking due to the momenta
of mesons.
Hence, it is useful to verify that this chiral invariant 
Lagrangian will lead eventually to the same results as those
obtained from the Lagrangians without derivatives in the fields.  
We also perform self-energy calculations using this $\bm{8}^s$
chirally symmetric Lagrangian, Eq.~(\ref{eq:chiralLag}).

\subsection{Explicit SU(3) breaking term}\label{subsec:massLag}

In this section, we consider the SU(3) breaking interaction
term within the context of chiral Lagrangians. 
Without using derivatives in the fields, 
the only possible term is a mass term that violates both SU(3) and
chiral symmetry, but in the way demanded by the underlying QCD
Lagrangian~\cite{Ecker:1995gg,Pich:1995bw,Bernard:1995dp}.
The mass term appears through the combination
\begin{equation}
    S=\xi M \xi +\xi^{\dagger} M \xi^{\dagger}  ,
    \label{eq:Smass}
\end{equation}
with the mass matrix,
written in terms of the meson masses,
\begin{equation}
   M=
   \begin{pmatrix}
       m_{\pi}^2 & &  \\
       & m_{\pi}^2 &  \\
       & & 2m_{K}^2-m_{\pi}^2
   \end{pmatrix}  .
   \label{eq:massmatrix}
\end{equation}
Then it leads to the Lagrangian
\begin{equation}
    \mathcal{L}^M= \frac{g^M}{2f} 
    \bar{P}_{ijk} \epsilon^{lmk}S_l{}^i B_m{}^j +
   \text{ h.c.} 
     ,     \label{eq:massLag}
\end{equation}
In the expansion of $S$, we have two meson fields
with the structure
\begin{equation}
    S^{(2)}=-\frac{1}{2f^2}(2\phi M \phi + \phi \phi M +M\phi\phi) . 
\end{equation}
Substituting $S^{(2)}$ for $S$ in Eq.~\eqref{eq:massLag}, 
we obtain the desired mass Lagrangian.

\section{Self-energies}\label{sec:Self}

\subsection{Two-meson loops}

The antidecuplet self-energies deduced from one of 
the interaction Lagrangians can be obtained by
\begin{equation}
    \begin{split}
        \Sigma^{(j)}_{P}(p^0)
        = &\sum_{B,m_1,m_2} \left( F^{(j)}
        C^{(j)}_{P,B,m_1,m_2} \right)  \\
        &\times I^{(j)}(p^0;B,m_1,m_2) 
        \left(F^{(j)} C^{(j)}_{P,B,m_1,m_2} \right) \ ,
    \end{split}
    \label{eq:selfenergy}
\end{equation}
where the index $j$ stands for $8s$, $8a$, $27$, $\chi$ and $M$
for corresponding Lagrangians~\eqref{eq:8sLag}, \eqref{eq:8aLag},
\eqref{eq:27Lag}, \eqref{eq:chiralLag} and \eqref{eq:massLag};
$P$ denotes the antidecuplet states
$P=\Theta_{\overline{10}}$, $N_{\overline{10}}$, $\Sigma_{\overline{10}}$ 
and $\Xi_{\overline{10}}$;
the argument $p^0$ is the energy of the antidecuplet baryon;
and the factors $F^{(j)}$ are
\begin{equation}
    \begin{split}
        F^{8s} &= \frac{g^{8s}}{2f}  , \quad 
        F^{8a} = \frac{g^{8a}}{4f^2}  , \\
        F^{27} = \frac{g^{27}}{2f}  ,& \quad
        F^{\chi} = \frac{g^{\chi}}{2f} , \quad
        F^{M} = \frac{g^{M}}{2f}  .
    \end{split}
    \label{eq:Ffactor}
\end{equation}
In Eq.~\eqref{eq:selfenergy},
$C^{(j)}_{P,B,m_1,m_2}$ are SU(3) coefficients that come
directly from the Lagrangians when evaluating the different matrix 
elements. We compile the results in Appendix~\ref{sec:Coef}.

\begin{figure}[tbp]
    \centering
    \includegraphics[width=7.5cm,clip]{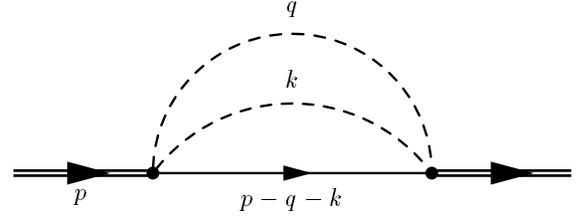}
    \caption{
    Self-energy of baryon antidecuplet caused by
    two-meson
    cloud.}
    \label{fig:loop2meson}
\end{figure}

The function $I^{(j)}(p^0;B,m_1,m_2)$
of argument $p^0$ (the energy of the assumed state of the antidecuplet
at rest)
is the two-loop integral with two mesons and one baryon as shown in
Fig.~\ref{fig:loop2meson}.
\begin{align}
    I^{(j)}(p^0;B,m_1,m_2)
    =& -
    \int \frac{d^4k}{(2\pi)^4}
    \int \frac{d^4q}{(2\pi)^4} \nonumber \\
    &\times|t^{(j)}|^2
    \frac{1}{k^2-m_1^2+i\epsilon}
    \frac{1}{q^2-m_2^2+i\epsilon} \nonumber \\
    &\times \frac{M}{E} 
    \frac{1}{p^0-k^0-q^0-E+i\epsilon}  ,
    \label{eq:loop2meson}
\end{align}
where
\begin{align}
    |t^{(j)}|^2 =& 1 \quad
    \text{for} \quad j=8s,27,M ,
    \label{eq:scalamp} \\
    |t^{\chi}|^2 =& \frac{(\omega_1\omega_2-\bm{k}\cdot\bm{q})^2}{4f^4} 
    , \label{eq:chiralamp} \\
    |t^{8a}|^2 =& 
    \frac{1}{2M}
    \Bigl\{
    (E+M)(\omega_1-\omega_2)^2 
    +2(|\bm{k}|^2-|\bm{q}|^2)(\omega_1-\omega_2)
    \nonumber \\
    &+(E-M)(\bm{k}-\bm{q})^2
    \Bigr\} , \label{eq:vecamp} \\
    E=&
    \sqrt{M^2+(\bm{k}+\bm{q})^2}  , \nonumber \\
    \omega_1=&\sqrt{m_1^2+\bm{k}^2}  , \quad
    \omega_2=\sqrt{m_2^2+\bm{q}^2}  . \nonumber
\end{align}
In these expressions, $M$ and $m_i$ are the masses of a baryon
and mesons.
The more complicated integrand in $|t^{(8a)}|^2$ arises because of the
$\bar{u} \gamma^\mu(k-q)_\mu u$ factor
when one derivative is included as in Eq.~\eqref{eq:8aLag}.
We neglect the negative-energy intermediate baryon propagator 
as this is suppressed by a further power of $q/M$,
leading only to a small relativistic correction.
The $k^0$ and $q^0$ integrations of Eq.~\eqref{eq:loop2meson}
are easily carried out, and we obtain
\begin{align}
    I^{(j)}(p^0;B,m_1,m_2)
    =& 
    \int \frac{d^3k}{(2\pi)^3}
    \int \frac{d^3q}{(2\pi)^3}
    |t^{(j)}|^2
    \frac{1}{2\omega_1}
    \frac{1}{2\omega_2} \nonumber \\
    &\times\frac{M}{E} 
    \frac{1}{p^0-\omega_1-\omega_2-E+i\epsilon} ,
    \label{eq:loop2meson2}
\end{align}
The real part of this integral is divergent.
We regularize it with a cutoff $\Lambda$ in the three momentum
on $\bm{k}$ and $\bm{q}$, which is a parameter of the calculation
and its value must be somewhat
larger than the scale of the typical pion momenta. 
On the other hand,
we use low-energy Lagrangians with one or two derivatives at most,
and thus the cutoff should not be too large;
otherwise, terms with more derivatives could become relevant.
In this work we will take $\Lambda$ in the range 700 -- 800 MeV,
roughly the order of magnitude of the cutoff used 
to regularize the meson-baryon
loops in the study of the $\bar{K}N$ interaction~\cite{Oset:1998it}.
With the $\mathcal{L}^{\chi}$ of Sec.~\ref{subsec:chiralLag},
the cutoff is smaller in order to
reproduce analogous results to those with the $8s$ Lagrangian.

The imaginary part of the diagram 
represents the decay width, in accordance with the optical theorem. 
The total decay width of a member of the antidecuplet
to any $BMM$ states is given by
\begin{equation}
    \Gamma^{(j)}_P(p^0)
    =-2\im \Sigma^{(j)}_P(p^0)  ,
    \label{eq:totalwidth}
\end{equation}
while the partial decay width to a particular channel is given by
\begin{align}
    \Gamma^{(j)}_P(p^0;B,m_1,m_2) 
    =&-2\im\left(F^{(j)}C^{(j)}_{P,B,m_1,m_2}\right) \nonumber \\
    &\times I^{(j)}(p^0;B,m_1,m_2)
    \left( F^{(j)}C^{(j)}_{P,B,m_1,m_2} \right) ,
    \label{eq:partialwidth}
\end{align}

As an example, let us give in detail the contribution from
$\mathcal{L}^{8a}$ to the $\Theta_{\overline{10}}$ self-energy
\begin{align}
    \Sigma^{8a}_{\Theta}(p^0)
    = &\left(F^{8a}\right)^2 
    [18I^{8a}(p^0;N,K,\pi) \nonumber \\
    &+18I^{8a}(p^0;N,K,\eta)] ,
    \label{eq:example1}
\end{align}
and the contribution from $\mathcal{L}^{8s}$ to the $\Xi_{\overline{10}}$
self-energy
\begin{align}
    \Sigma^{8s}_{\Xi}(p^0)
    =& \left(F^{8s}\right)^2 
    [9I^{8s}(p^0;\Sigma,\bar{K},\pi)
    +I^{8s}(p^0;\Sigma,\bar{K},\eta)  \nonumber \\
    &+6I^{8s}(p^0;\Xi,\bar{K},K)
    +4I^{8s}(p^0;\Xi,\pi,\eta)] .
    \label{eq:example2}
\end{align}
The expression for all cases can be derived from 
Tables~IX--XII in Appendix~\ref{sec:Formulae}.

In Eq.~(\ref{eq:selfenergy}), we gave a contribution to the 
self-energy from one interaction Lagrangian $\mathcal{L}^{(j)}$.  
For the total self-energy, the sum should be taken over the five 
interactions ($j = 8s, 8a, 27, \chi$ and $M$) at each vertex.  
This means that at each vertex function, we should make the 
replacement as 
$(F^{(j)} C^{(j)}_{P,B,m_1,m_2}|t^{(j)}| )
\to 
(|\sum_j F^{(j)} C^{(j)}_{P,B,m_1,m_2}t^{(j)}|)$.  
We shall, however, not take into account interference between the $8a$
term and the others because of the $p$-wave nature of the term.

\subsection{Inclusion of the $\rho$ meson}

\begin{figure}[tbp]
    \centering
    \includegraphics[width=7.5cm,clip]{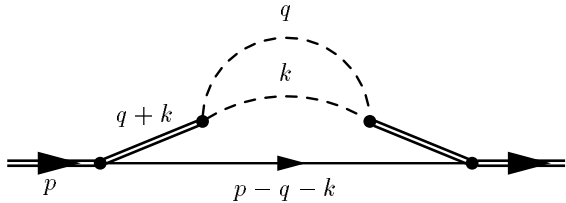}
    \caption{
    Self-energy of baryon antidecuplet caused by
    two-meson
    cloud with vector meson propagators.}
    \label{fig:loopvec}
\end{figure}

It is known that $N(1710)\to N\pi\pi(p$ wave) occurs through 
the $N\rho$ decay.
In order to keep close to the experimental
information, we shall also assume that the pair of mesons in the
$8a$ case reconstruct a  vector meson.
Hence, we  
replace the contact interaction of the $\mathcal{L}^{8a}$ to account 
for the vector meson propagator (Fig.~\ref{fig:loopvec})
and include the factor
\begin{equation}
    \frac{m_v^2}{(q+k)^2-m_v^2}  ,
    \label{eq:vecprop}
\end{equation}
in each $P\to BMM$ vertex.
The consideration of these contributions needs extra work
on the loop integrals since we introduce new poles.
The imaginary part of the integrals (associated to placing on-shell
the $BMM$ intermediate states) can be easily accounted for by
multiplying the integrand of Eq.~\eqref{eq:loop2meson} by
\begin{equation}
    \left|
    \frac{m_v^2}{(q+k)^2-m_v^2+im_{v}\Gamma(q+k)}
    \right|^2 ,
    \label{eq:imagfac}
\end{equation}
where $\Gamma(q+k)$ accounts for the width of the vector meson
($\rho$ or $K^*$ depending on the $MM$) incorporating the energy
dependence through the factor $(P(q+k)/P_{on})^3$ multiplied to the
nominal width, with $P(q+k)$ the relative three momentum of the
mesons in the decay of the vector meson in the rest frame and
$P_{on}=P((M_v,\bm{0}))$.

For the real part, one must sort out the poles of the vector meson
and the intermediate $BMM$ state, which is technically implemented 
by means of the integral
\begin{equation}
    \begin{split} 
	&\re \left\{ I^{8a}(p^0;B,m_1,m_2) \right\} \\
	=&
	-m_v^4
	\frac{\partial}{\partial (m_v^2)}
	\text{P.V.}
	\int \frac{d^4q}{(2\pi)^4}
	\int \frac{d^4k}{(2\pi)^4}
	|t^{8a}|^2\\
	&\times
	\frac{1}{k^2-m_1^2+i\epsilon}
	\frac{1}{q^2-m_2^2+i\epsilon}\\
	&\times
	\frac{1}{(k+q)^2-m_v^2+i\epsilon}\\
	&\times
	\frac{M}{E}
	\frac{1}{p^0-k^0-q^0-E+i\epsilon}  ,
    \end{split}
    \label{eq:loopvec1}
\end{equation}
where P.V. stands for the principal value.
Here, we neglected the width of the vector meson,
which does not play much of a role in the off-shell regions of
integrations.
The $k^0$ and $q^0$ integrations can be performed
analytically, and one obtains the simple expression
\begin{equation}
    \begin{split} 
	&\re \left\{ I^{8a}(p^0;B,m_1,m_2) \right\} \\
	=&m_v^4 
	\frac{\partial}{\partial (m_v^2)}
	\text{P.V.}\int \frac{d^3k}{(2\pi)^3}
	\int \frac{d^3q}{(2\pi)^3} |t^{8a}|^2\\
	&\times
	\frac{1}{2\omega_1}
	\frac{1}{2\omega_2}
	\frac{1}{\omega_v}\frac{1}{\omega_v+\omega_1+\omega_2}\\
	&\times
	\frac{1}
	{p_0-\omega_v-E+i\epsilon}\\
	&\times
	\frac{1}
	{p_0-\omega_1-\omega_2-E+i\epsilon}
	\frac{M}{E} \\
	&\times (\omega_1+\omega_2+\omega_v-p_0+E)  ,
    \end{split}
    \label{eq:loopvec2}
\end{equation}
where $\omega_v$ is the on-shell energy of the vector meson.

\section{Numerical examples}\label{sec:Results}

Next we present some numerical results that illustrate 
the antidecuplet mass shifts and decay widths to three-body channels.
One of the most exciting aspects in the antidecuplet
is that the $\Theta^+$ is located about
$30$ MeV below the $NK\pi$ threshold.  
Hence, it cannot decay into this or any other
$BMM$ channels to which it couples.
For the interaction Lagrangians, we obtain the $g^{(j)}$
coefficients from the experimentally allowed 
decay amplitudes of the $N(1710)$.
We give several examples that illustrate the
general behavior of the two-meson cloud, common to the Lagrangians
described in previous sections.

Before studying each of the Lagrangians, let us 
recall that the mass splitting of the antidecuplet
has a contribution which follows the GMO rule, 
and it would be originated by the
difference of the masses of the constituent quarks
and their correlations.
To this, we add the splitting coming from the real part
of the self-energy due to the meson cloud that we are studying.
Thus, the masses of the antidecuplet are approximately given by
\begin{equation}
    \begin{split}
        M_{\Theta_{\overline{10}}}
        =& M_0 + \re \Sigma_{\Theta_{\overline{10}}} , \\
        M_{N_{\overline{10}}}
        =& M_0 + \re \Sigma_{N_{\overline{10}}} + \Delta , \\
        M_{\Sigma_{\overline{10}}}
        =& M_0 + \re \Sigma_{\Sigma_{\overline{10}}} + 2\Delta , \\
        M_{\Xi_{\overline{10}}}
        =& M_0 + \re \Sigma_{\Xi_{\overline{10}}} + 3\Delta ,
    \end{split}
    \label{eq:massbar10}
\end{equation}
where $M_0$ is the bare mass of the antidecuplet
and $\Delta$ is the GMO mass splitting,
part of which simply comes from the difference of the constituent quark
masses.
In the constituent quark model, $\Delta$ is related to the difference
between the constituent masses of $u$, $d$ and $s$ quarks,
$3\Delta = \langle m_s-m_{u,d}\rangle_{\text{baryon}}$. 
Certainly, quark correlations can also contribute to the experimental
value of $\Delta$.

The difference between the light and strange quark masses has been 
obtained, for example from hyperfine splittings,
in Ref.~\cite{lastlipkin},
\begin{equation}
    \begin{split}
        \la m_s-m_u\ra_{\rm meson}
        &= \frac{3(M_{K^*}-M_\rho)+(M_K-M_\pi)}{4} \\
        &\simeq 180 \text{ MeV} ,
    \end{split}
\end{equation}
whereas for baryons 
\begin{equation}
    \begin{split}
        \la m_s-m_u\ra_{\rm baryon}
        =& M_\Lambda - M_ N 
        \simeq 177 \text{ MeV} , \\
        \la m_s-m_u\ra_{\rm baryon}
        =&\frac{M_N+M_\Delta}{6} \left(
        \frac{M_\Delta-M_N}{M_{\Sigma^*}-M_\Sigma}-1 \right) \\
        \simeq & 190 \text{ MeV} .
    \end{split}
    \label{eq:baryonmass}
\end{equation}
But other differences like $M_{K^*}-M_\rho$,
$M_\Xi-M_N$ or $M_\Sigma-M_N$ suggest a wider range, from 
122 to 190 and 250 MeV, respectively.
As we will see, the values of $3\Delta$ 
needed within this work are of this order of
magnitude but somewhat larger,  
leaving room for extra quark correlations effects.

\subsection{Antidecuplet mass shift with ${\mathcal L}^{8s}$
and ${\mathcal L}^{8a}$ }

To fix the couplings of the Lagrangians, we start by
taking $\mathcal{L}^{8s}$ and $\mathcal{L}^{8a}$ defined above
and adjusting the coupling constants to obtain the partial decay widths
of the $N(1710)$ to $N\pi\pi(s$ wave, isoscalar) and 
$N\rho\to N\pi\pi(p$ wave, isovector)
respectively. These are controlled by the imaginary part of the 
self-energies~\eqref{eq:partialwidth}, which are finite and
independent of the cutoff. 
The central values in the Particle Data Group (PDG)~\cite{Eidelman:2004wy} are
\begin{equation}
    \begin{split}
        \Gamma(N\pi\pi, s\text{ wave})
        &=25 \text{ MeV} , \\
        \Gamma(N\pi\pi, p\text{ wave})
        &=15 \text{ MeV} ,
    \end{split}
    \label{eq:Nstardecay}
\end{equation}
and the uncertainties (counting those of the branching ratio and the 
total width) can be a large fraction of these numbers.

\begin{figure}[tbp]
    \centering
    \includegraphics[width=7.5cm,clip]{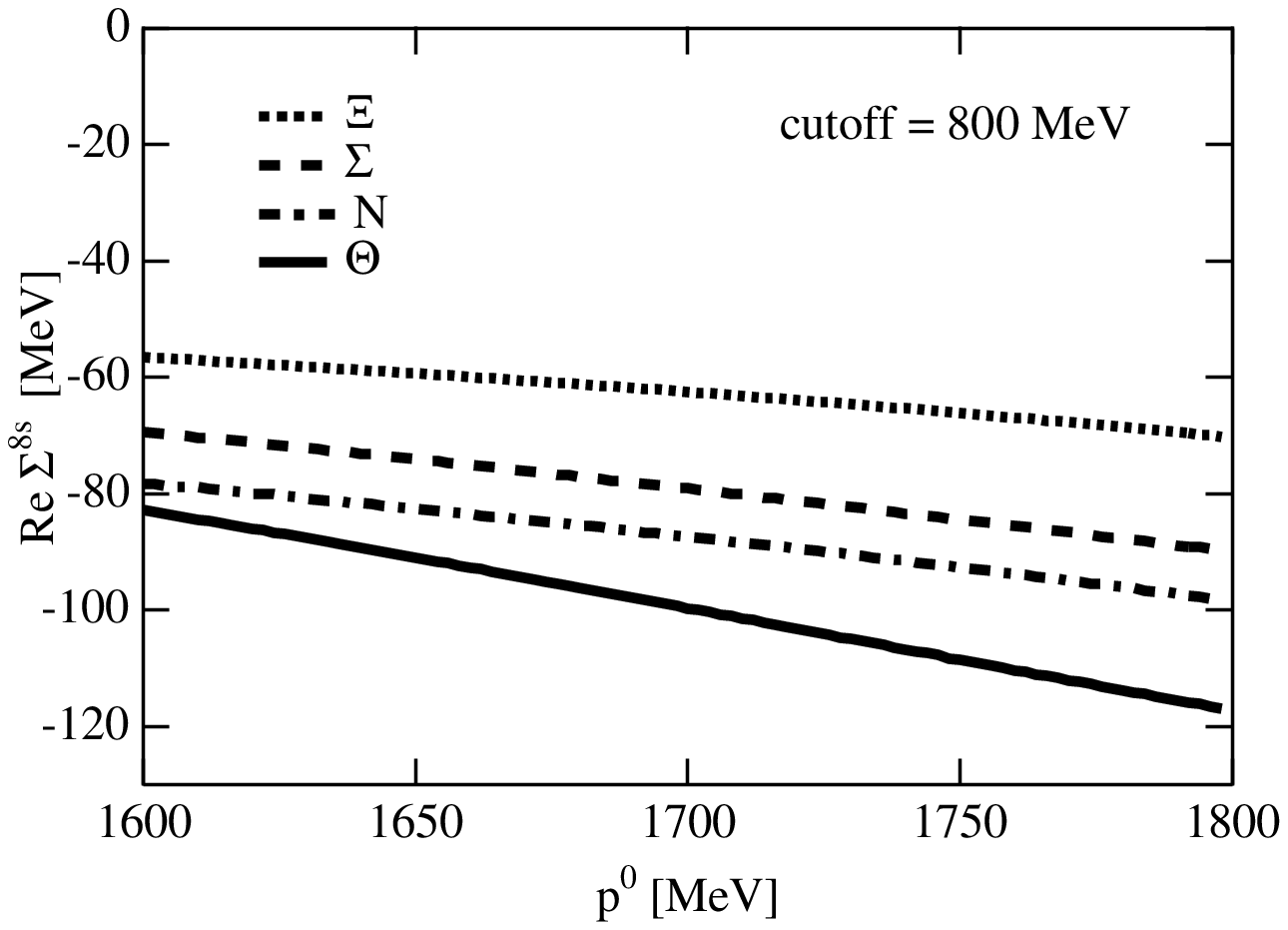}
    \caption{
    Mass shifts of baryon antidecuplet $(\re \Sigma_P)$ due to
    two-meson cloud from $\mathcal{L}^{8s}$ with cutoff $= 800$ MeV;
    $p^0$ dependence.}
    \label{fig:pzerodep}
\end{figure}

A fit to these central values gives us
\begin{equation}
    g^{8s} = 1.9 , \quad
    g^{8a} = 0.32  .
    \label{eq:gfactors}
\end{equation}
With these couplings, we calculate the real part of the
self-energies for all the antidecuplet.
For the bare antidecuplet mass $p^0$ as input,
we take an average value of $p^0=1700$ MeV.
We also performed a calculation with different values of $p^0$ 
and found that the results have the same qualitative trend,
but the depth of the binding varies.
To estimate the binding, we show the mass
shift from the $\mathcal{L}^{8s}$ with respect to $p^0$ in 
Fig.~\ref{fig:pzerodep}.
We see that, independently of the values of $p^0$,
all the self-energies are attractive,
and that the interaction is more attractive  the larger
the strangeness; hence, the $\Theta_{\overline{10}}$ is always more bound.

\begin{figure*}[tbp]
    \centering
    \includegraphics[width=7.5cm,clip]{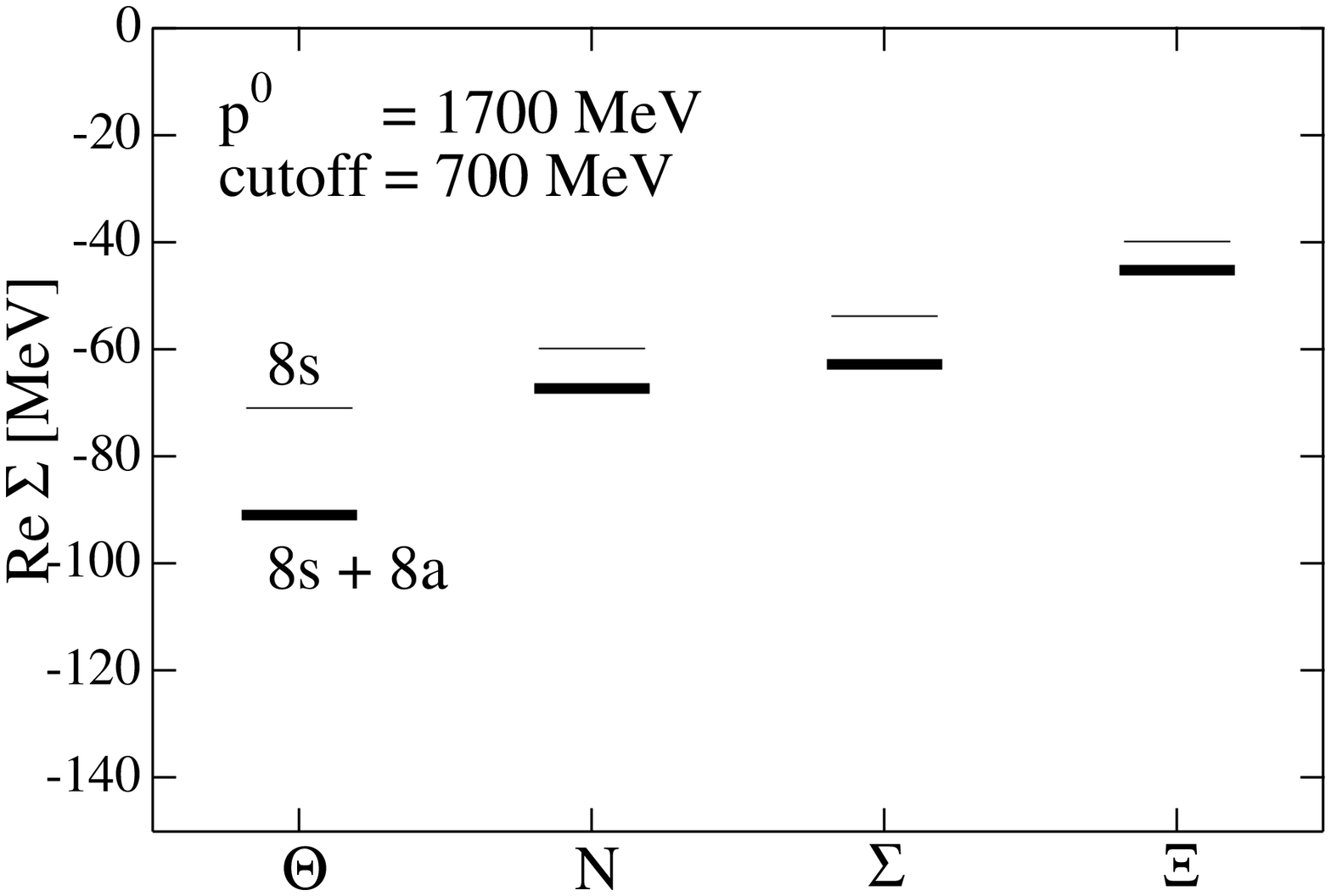}
    \includegraphics[width=7.5cm,clip]{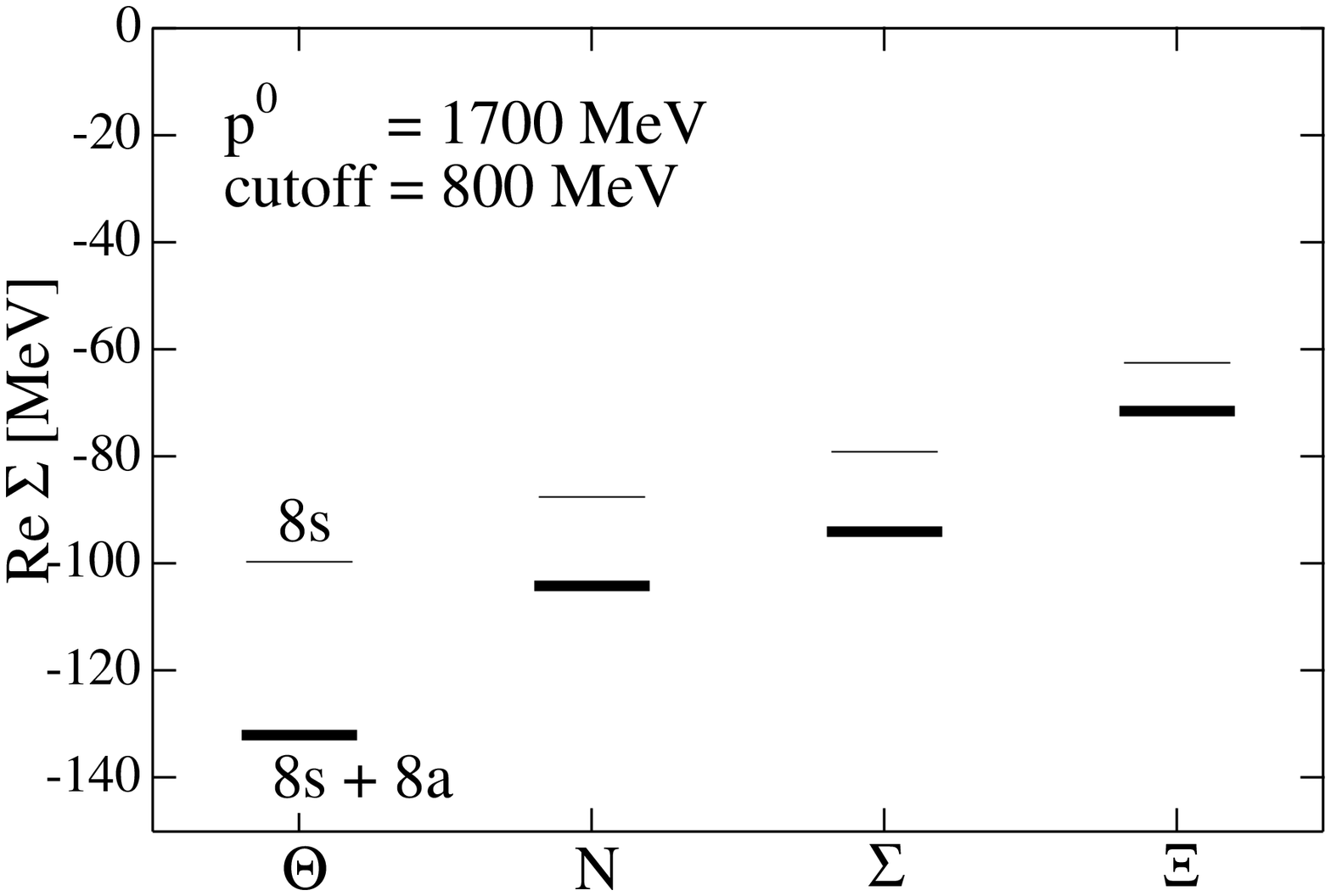}
    \caption{
    Mass shifts of baryon antidecuplet $(\re \Sigma_P)$  due to
    two-meson cloud with $p^0 = 1700$ MeV
    at two cutoff values.
    Thin lines represent the results from contributions
    from $\mathcal{L}^{8s}$, and thick lines 
    denote the total contribution
    with $\mathcal{L}^{8s}$ and $\mathcal{L}^{8a}$.}
    \label{fig:totresult}
\end{figure*}

In Fig.~\ref{fig:totresult}
we show the results for the
contributions from $\mathcal{L}^{8s}$ 
and total contributions of  $\mathcal{L}^{8a}$
and  $\mathcal{L}^{8s}$,
with $p^0=1700$ MeV and cutoffs  
$700$ and $800$ MeV.
The numerical values of the mass shifts are displayed in
Table~\ref{tbl:totresult}.
We see that $\mathcal{L}^{8s}$ provides more binding than 
$\mathcal{L}^{8a}$ for the same cutoff.
The total binding for $\Theta_{\overline{10}}$
ranges from 90 to about 130 MeV, depending on the cutoff.
The splitting between the $\Theta_{\overline{10}}$ and 
$\Xi_{\overline{10}}$ states
is about 45 MeV for a cutoff of 700 MeV and 60 MeV for
a cutoff of 800 MeV. Since the experimental splitting
is 320 MeV for the $\Theta(1540)$ and $\Xi(1860)$,
the splitting provided by the two-meson cloud
is on the order of 20\% of the experimental one.

\begin{table}[btp]
    \centering
    \caption{Mass shifts of baryon antidecuplet $(\re \Sigma_P)$ 
    due to two-meson cloud with $p^0=1700$ MeV and
    cutoffs 700 and 800 MeV.
    All values are shown in units of MeV.}
   \begin{ruledtabular}
        \begin{tabular}{rrrrrrr}
            & \multicolumn{3}{l}{cutoff 700 MeV} & 
            \multicolumn{3}{l}{cutoff 800 MeV}  \\
             & $\mathcal{L}^{8s}$ & $\mathcal{L}^{8a}$ & total
             & $\mathcal{L}^{8s}$ & $\mathcal{L}^{8a}$ & total  \\
	     \hline
            $\Theta_{\overline{10}}$ & $-71$ & $-20$ & $-91$
           &   $-100$ & $-32$ & $-132$ \\
            $N_{\overline{10}}$ & $-60$ & $-7$ & $-67$ 
           &  $-87$ & $-17$ & $-104$ \\
            $\Sigma_{\overline{10}}$ & $-54$ & $-9$ & $-63$ 
           &  $-79$ & $-15$ & $-94$ \\
            $\Xi_{\overline{10}}$ & $-40$ & $-5$ & $-45$ 
           &$-63$ & $-9$ & $-72$ 
        \end{tabular}
\label{tbl:totresult}
\end{ruledtabular}
\end{table}

We believe these magnitudes to be realistic (and hence
one of the reasons to settle for a cutoff)
based on the findings of Ref.~\cite{Llanes-Estrada:2003us}
that the meson-baryon interaction is insufficient to bind 
the $K\pi N$ system and that one has to increase the interaction by 
about a factor of 5 to have the three-particle system bound.
Indeed, had the nature of the $\Theta(1540)$ been that of the 
$K\pi N$ system, we would have obtained all the splitting from the
two-meson cloud.
There is, hence, a qualitative correlation between the moderate amount 
of the two-meson cloud contribution
claimed here and the difficulty to make
the stable $K\pi N$ system based alone on the $K\pi N$ dynamics.

Next we present the antidecuplet spectrum 
generated with the splitting obtained here.
We take the cutoff 800 MeV for reference. 
Inserting $M_{\Theta_{\overline{10}}}=1540$ MeV and  
$M_{\Xi_{\overline{10}}}=1860$ MeV in Eq.~\eqref{eq:massbar10},
together with our calculated self-energies, 
we obtain $M_0=1670$ MeV and $\Delta=87.5$ MeV, then
\begin{equation}
    \begin{split}
        M_{\Theta_{\overline{10}}}
        =& 1540\text{ MeV} \quad (\text{input}) , \\
        M_{N_{\overline{10}}}
        =& 1652\text{ MeV} , \\
        M_{\Sigma_{\overline{10}}}
        =& 1749\text{ MeV} , \\
        M_{\Xi_{\overline{10}}}
        =& 1860\text{ MeV} \quad (\text{input}) .
    \end{split}
    \label{eq:massbar10res}
\end{equation}
The value $3\Delta\sim 260$ MeV
is fairly reasonable for our estimate purposes. 
It would indicate, however, that about 30 MeV of $\Delta$, 
above the 60 MeV coming from the constituent quark consideration, 
would come from quark correlations.  
The large $\Theta_{\overline{10}}$ binding with
respect to that of
the $N_{\overline{10}}$ state is responsible for the new value
$M_{N_{\overline{10}}}=1652$ MeV, slightly higher than the value we
would obtain from an exact GMO rule splitting (1646 MeV),
but still far from the 1710 MeV resonance
we have assumed for the antidecuplet.
As discussed in the introduction, a necessary mixture of an octet
representation with the antidecuplet could bring the mass close to
that of the $N(1710)$, although the possibility of having a new $N^*$
resonance belonging to the antidecuplet cannot be ruled 
out~\cite{Arndt:2003ga}.

\subsection{Antidecuplet decay widths from ${\mathcal L}^{8s}$ and
${\mathcal L}^{8a}$}

Now we show the partial decay width obtained
according to Eq.~\eqref{eq:partialwidth}. 
As already mentioned, $\Theta(1540)$ has no $BMM$ channel to decay.
Among all decay channels,
the $N(1710)$ decays broadly into $N\pi\pi$, and it can also decay
into $N\pi\eta$.
The $\Sigma(1770)$ can decay
into $N\bar{K}\pi$, $\Lambda\pi\pi$, and $\Sigma\pi\pi$, and the
$\Xi(1860)$ into $\Sigma\bar{K}\pi$ and $\Xi\pi\pi$,
because of the threshold energies of $BMM$ channels.

To calculate the decay, since the phase space is essential for the
imaginary part, we take the observed masses,
\begin{equation}
    M_{N_{\overline{10}}} = 1710  , \quad 
    M_{\Sigma_{\overline{10}}} = 1770 , \quad 
    M_{\Xi_{\overline{10}}} = 1860 .
    \label{eq:decayinput}
\end{equation}
The results appear in Table~\ref{tbl:decay}.
We can see that the widths are not very large for all channels.
Among them, we obtain the 
partial decay widths of the $\Sigma(1770)$
into $\Sigma\pi\pi$  and $N \bar{K} \pi$.
When compared with the experimental data,
indeed, the $\Sigma(1770)$ would have a total width
into two-meson and baryon of about 24 MeV, which is well within the
total width of the $\Sigma(1770)$ of about 70 
MeV~\cite{Eidelman:2004wy}.
As to the $\Xi(1860)$ resonance, we obtain a total width of about
$2$ MeV, which is certainly compatible with the total width less smaller
than 18 MeV claimed by the NA49 collaboration~\cite{Alt:2003vb}.
Detailed information of the partial decay widths of these
resonances to three-body channels will give us more understanding
of the $PBMM$ interaction.

\begin{table}[btp]
    \centering
    \caption{Partial decay widths for the allowed channels and
    total width for any $BMM$ channel, at the
    masses of the antidecuplet members. All values are 
    in MeV.}
    \begin{ruledtabular}
    \begin{tabular}{llll}
    Decay  widths (MeV) & $\Gamma^{(8s)}$ & $\Gamma^{(8a)}$
    & $\Gamma^{tot}_{BMM}$ \\ \hline
    $N(1710)\to N\pi\pi$ (inputs) & 25 & 15 & 40 \\
    $N(1710)\to N\eta\pi$ & \phantom{ }0.58 & - &  \vspace{0.1cm} \\ 
    $\Sigma(1770)\to N\bar{K}\pi$ & \phantom{ }4.7 
    & \phantom{ }6.0 & 24 \\
    $\Sigma(1770)\to \Sigma\pi\pi$ & 10 & \phantom{ }0.62  &\\
    $\Sigma(1770)\to \Lambda\pi\pi$ & - & \phantom{ }2.9 &  \vspace{0.1cm}
    \\ 
    $\Xi(1860)\to \Sigma \bar{K}\pi$ & \phantom{ }0.57 
    & \phantom{ }0.46 &  \phantom{ }2.1 \\
    $\Xi(1860)\to \Xi \pi \pi$ & - & \phantom{ }1.1 & \\
    \end{tabular}
    \end{ruledtabular}
    \label{tbl:decay}
\end{table}

\subsection{Mass shifts and decay widths from ${\mathcal L}^{\chi}$}

Here we show the results for the interaction Lagrangian
given in Sec.~\ref{subsec:chiralLag}, namely
the two-meson coupling derived from the chiral symmetric Lagrangian
$\mathcal{L}^{\chi}$.
We fix the coupling constant $g^{\chi}$ from
the $N(1710)$ decay to $N\pi\pi(s$ wave, isoscalar),
and we find $g^{\chi} = 0.218$.
Then the antidecuplet mass shifts and decay widths are calculated.
However, for the mass shifts,
we obtain binding energies that are too large---on the order of several hundreds
MeV with cutoffs around
700-800 MeV---because the loop integral is more divergent than
the previous $\mathcal{L}^{8s}$ case.
To reach some reasonable results, we decrease the cutoff,
and find that $\Lambda=525$ MeV would give  mass shifts
similar to those of $\mathcal{L}^{8s}$ without derivatives.

We compare the mass shifts of $\mathcal{L}^{\chi}$
with cutoff 525 MeV and $\mathcal{L}^{8s}$ with 800 MeV
in Table~\ref{tbl:chiralresult}.
The decay widths obtained from these Lagrangians are given in 
Table~\ref{tbl:chiraldecay}.
As expected from the fact that the $C^{(j)}$ coefficients of 
two Lagrangians are identical, 
we obtain almost the same mass shifts for
$\mathcal{L}^{\chi}$ and $\mathcal{L}^{8s}$ by properly 
adjusting
the cutoffs.
The decay widths are considered to be in fair agreement
qualitatively, when considering
that the values span two orders of magnitude.
Some quantitative differences would come from the
SU(3) breaking in the meson momenta appearing in the
$\mathcal{L}^{\chi}$ loop, and they are regarded as the uncertainty
in our analysis.

\begin{table}[btp]
    \centering
    \caption{Mass shifts of baryon antidecuplet $(\re \Sigma_P)$ 
    due to two-meson cloud with 800 MeV cutoff
    for $\mathcal{L}^{8s}$ and 525 MeV for
    $\mathcal{L}^{\chi}$.
    All values are in MeV.}
    \begin{ruledtabular}
        \begin{tabular}{lrr}
            Mass shifts (MeV) 
            & $\mathcal{L}^{8s}$ & $\mathcal{L}^{\chi}$  \\ \hline
            $\Theta_{\overline{10}}$ & $-100$ & $-99$ \\
            $N_{\overline{10}}$ & $-87$ & $-83$ \\
            $\Sigma_{\overline{10}}$ & $-79$ & $-70$ \\
            $\Xi_{\overline{10}}$ & $-63$ & $-57$ 
        \end{tabular}
\label{tbl:chiralresult}
\end{ruledtabular}
\end{table} 

\begin{table}[btp]
    \centering
    \caption{Partial decay widths for the allowed channels
    with $\mathcal{L}^{8s}$ and $\mathcal{L}^{\chi}$, at the
    masses of the antidecuplet members. All values are
    in MeV.}
    \begin{ruledtabular}
    \begin{tabular}{llll}
    Decay widths [MeV] & $\Gamma^{(8s)}$ & $\Gamma^{(\chi)}$ \\ \hline
    $N(1710)\to N\pi\pi$ (input) & 25 & 25 \\
    $N(1710)\to N\eta\pi$ & \phantom{ }0.58 & 0.32 \\
    $\Sigma(1770)\to N\bar{K}\pi$ & \phantom{ }4.7 
    & \phantom{ }4.5 \\
    $\Sigma(1770)\to \Sigma\pi\pi$ & 10 & \phantom{}3.6 \\
    $\Xi(1860)\to \Sigma \bar{K}\pi$ & \phantom{ }0.57 
    & \phantom{ }0.40 \\
    \end{tabular}
    \end{ruledtabular}
    \label{tbl:chiraldecay}
\end{table}

\subsection{Effects of $\mathcal{L}^{27}$ and ${\mathcal L}^M$}
\label{subsec:27M}

Next we draw our attention to the $\mathcal{L}^{27}$ 
and $\mathcal{L}^{M}$ Lagrangians, that we have not yet used.
First note that
it is unrealistic to make these Lagrangians
solely responsible for the N(1710) decay width
into $N\pi\pi(s$ wave) channel.
This would lead to some unphysical results such as
very large decay widths of the $\Xi_{\overline{10}}$ into 
$BMM$ channels, or a large binding energy of several hundred MeV.
Hence, combined with the analyses in the previous section,
this fact would justify
the approach followed in Ref.~\cite{Cabrera:2004yg},
where only the $\mathcal{L}^{8s}$ and $\mathcal{L}^{8a}$ terms
are taken to study the $\Theta^+$ self-energies in a nuclear medium.
Thus, assuming that one cannot have a large fraction of these
Lagrangians, we will determine to what extent we can allow 
the contributions from $\mathcal{L}^{27}$ and $\mathcal{L}^M$.

\begin{figure*}[tbp]
    \centering
    \includegraphics[width=7.5cm,clip]{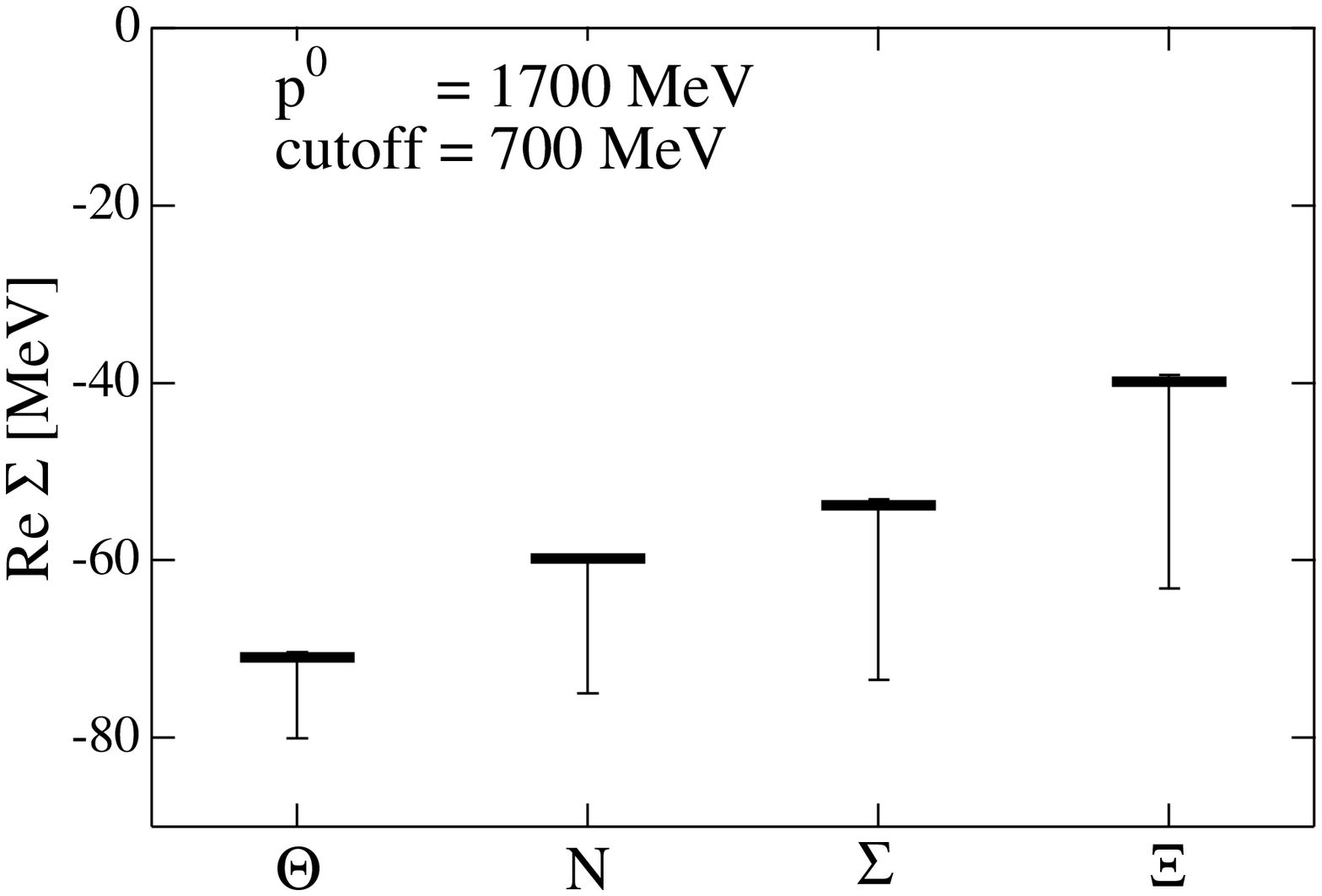}
    \includegraphics[width=7.5cm,clip]{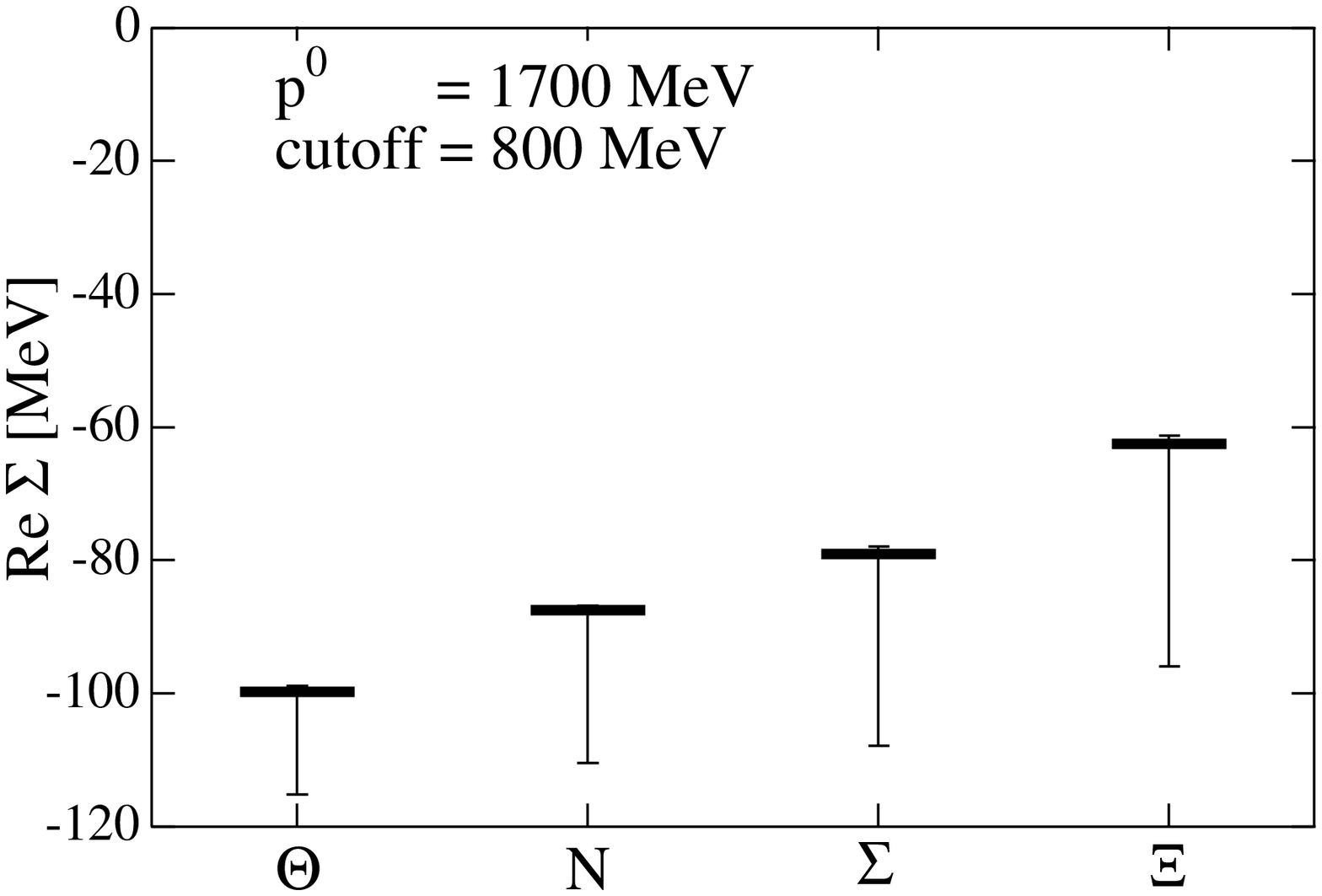}
    \caption{
    Mass shifts of baryon antidecuplet due to
    two-meson cloud
    from $\mathcal{L}^{8s}$ and possible $\mathcal{L}^{27}$
    contributions with two cutoff values.
    Horizontal bar: results with $\mathcal{L}^{8s}$ only.
    Vertical bar: band of values including 
    $\mathcal{L}^{27}$ in the range of the text.}
    \label{fig:27result}
\end{figure*}

We first pursue a model that mixes
$\mathcal{L}^{8s}$ and $\mathcal{L}^{27}$.
The coupling constants should be determined such that
the decay width of $N_{\overline{10}}\to N\pi\pi(s$-wave)
is unchanged.
According to Table~\ref{tbl:Coef2}, the $C^{(j)}_{B,m_1,m_2}$
coefficients for $N_{\overline{10}}\to N\pi\pi$ channels are
\begin{equation}
    \begin{split}
        C_{p\pi^0\pi^0}^{8s}&=\frac{1}{\sqrt{2}} \ , \quad
        C_{p\pi^0\pi^0}^{27}=-\frac{2\sqrt{2}}{5} \ , \\
        C_{p\pi^+\pi^-}^{8s}&=\sqrt{2} \ , \quad
        C_{p\pi^+\pi^-}^{27}=-\frac{4\sqrt{2}}{5} \  .
    \end{split}
    \nonumber
\end{equation}
To see the contribution from each Lagrangian clearly,
we set $g^{8s}=g^{27}=1.88$, and take the combination
\begin{equation}
    a\mathcal{L}^{8s}+b\mathcal{L}^{27} \ ,
    \quad
    b=-\frac{5}{4}(1-a)
    \label{eq:Lagcomb1}
\end{equation}
In this case,
\begin{equation}
    \begin{split}
        C_{p\pi^0\pi^0}^{8s+27}
        &=\frac{1}{\sqrt{2}}a
        -\frac{2\sqrt{2}}{5}\times\left(-\frac{5}{4}(1-a)\right)
        = \frac{1}{\sqrt{2}} \  , \\
        C_{p\pi^+\pi^-}^{8s+27}
        &=\sqrt{2}a-\frac{4\sqrt{2}}{5}
        \times\left(-\frac{5}{4}(1-a)\right) = \sqrt{2} \ ,
    \end{split}
    \nonumber
\end{equation}
and, therefore, we have the same $N(1710)\to N\pi\pi(s$ wave)
decay independent of $a$,
but different decays into other channels.
With this parametrization,
$a=1$ corresponds to the limit
where $\mathcal{L}^{27}$ is switched off,
while $a=0$ relates to the $\mathcal{L}^{27}$ contribution only.
We vary $a$ around $1$ and find that for 
$0.90 < a < 1.06$, the self-energy results are
acceptable on physical grounds.
If we exceed this range, the splitting of the different strangeness
states of the antidecuplet 
spoils agreement with the
GMO rule.
Taking this range of acceptable values of $a$ into account,
we find the results for the binding energies shown in
Fig.~\ref{fig:27result}.
As we see in the figure, $\mathcal{L}^{27}$ tends to
contribute to make the binding energy deeper.
A possible contribution from $\mathcal{L}^{27}$ would be
considered as a theoretical uncertainty in our analysis.

\begin{figure*}[tbp]
    \centering
    \includegraphics[width=7.5cm,clip]{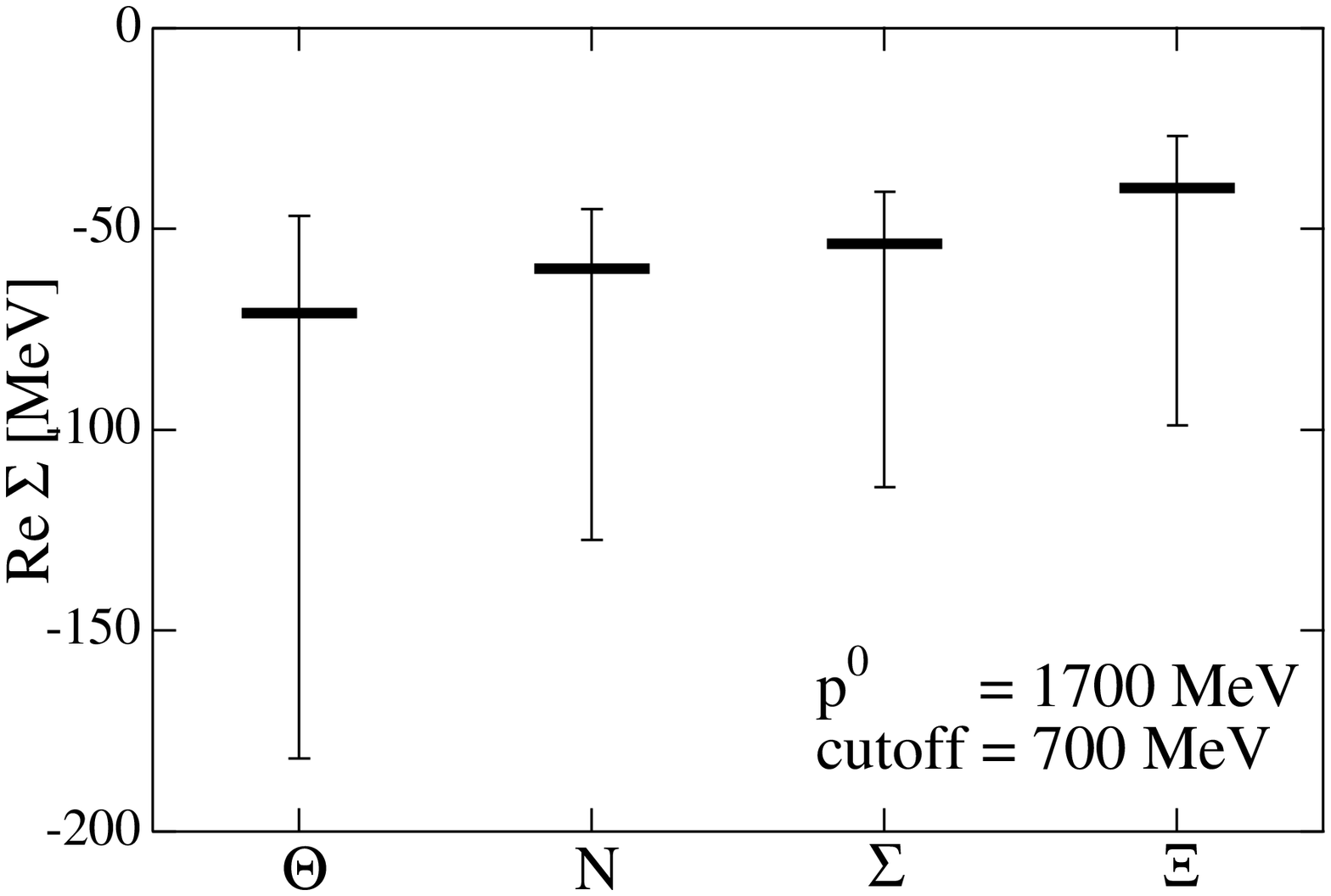}
    \includegraphics[width=7.5cm,clip]{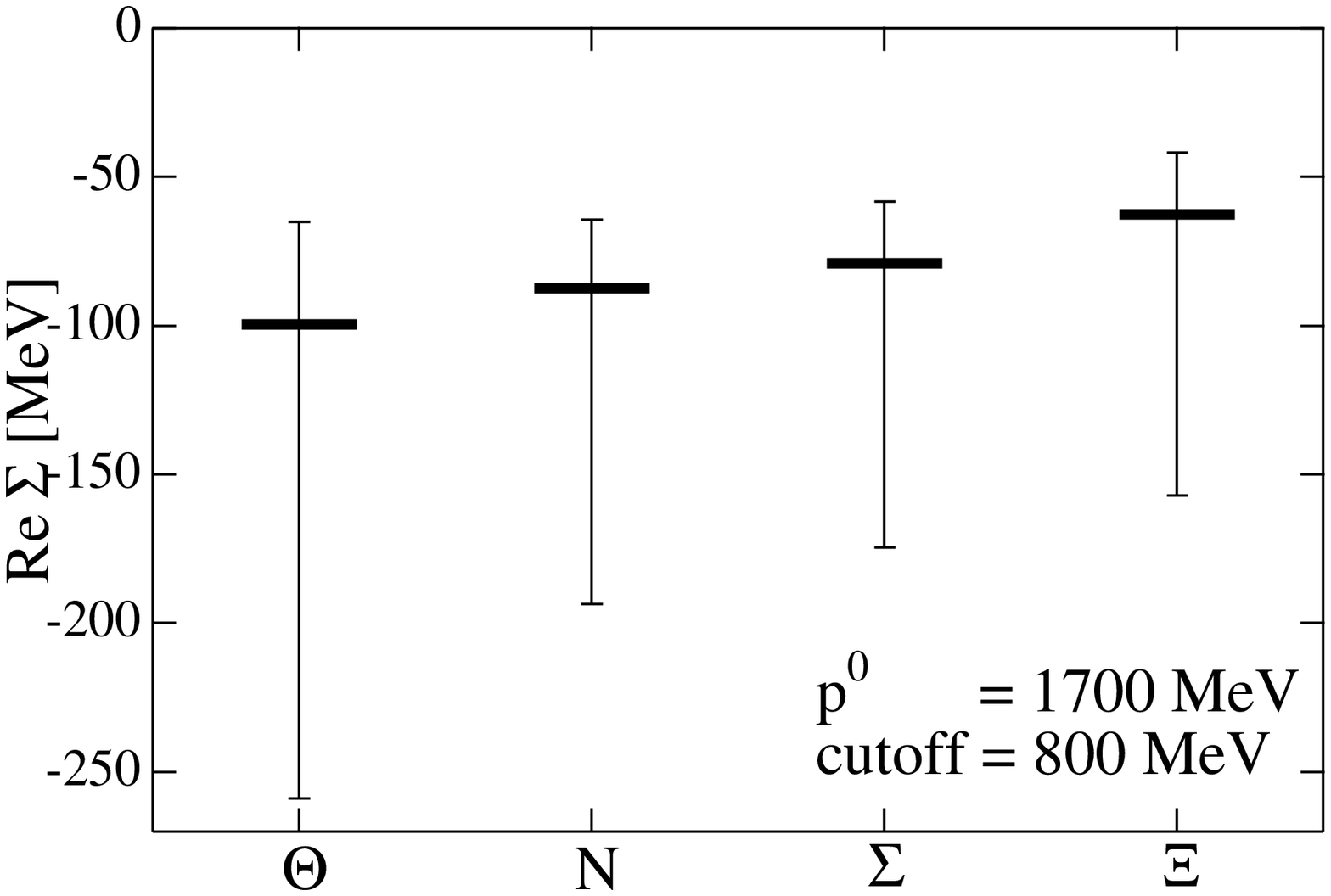}
    \caption{
    Mass shifts of baryon antidecuplet due to
    two-meson cloud
    from $\mathcal{L}^{8s}$ and possible $\mathcal{L}^{M}$
    contributions with two cutofff values. 
        Horizontal bar: results with $\mathcal{L}^{8s}$ only.
    Vertical bar: band of values including 
    $\mathcal{L}^{M}$ in the range of the text.}
   \label{fig:massresult}
\end{figure*}

Next we address the  $\mathcal{L}^M$ term.
Once again, as in the $\mathcal{L}^{27}$ case,
we set $g^{8s}=g^{M}=1.88$ and take
the combination
\begin{equation}
   a \mathcal{L}^{8s} + b\mathcal{L}^{M} \ , \quad
   b = \frac{f^2}{m_{\pi}^2}(1-a) \ ,
   \label{eq:Lagcomb2}
\end{equation}
in order to have the same  $N(1710)\to N\pi\pi(s$ wave).
In this case, we also see that the values of
$0.76 < a < 1.06$ are acceptable on physical grounds,
but larger deviations of $a$ again lead to undesired signs of the
splitting between members of the antidecuplet,
as well as to unacceptably large results of the binding energies.
Within this interval of coupling constant,
the results obtained for the binding energies of the antidecuplet
members are given in Fig.~\ref{fig:massresult}.
We observe that $\mathcal{L}^M$ also contributes to attractive
binding energy, and
the splitting of $\Theta_{\overline{10}}$ and 
$N_{\overline{10}}$ becomes large
compared with the other splittings.

\section{Discussion and conclusion}\label{sec:Summary}

The assumptions made throughout the paper and the uncertainties in
the experimental input make the nature of our analysis qualitative.
We assume that the $\Theta^+$ is
a $1/2^+$ state with $I=0$ and that it belongs to an antidecuplet.
In addition to these minimal assumptions, we consider that the $N(1710)$ 
also belongs to this same antidecuplet.
The meson cloud mechanism proposed in this work
leads, in all different cases studied, to the following conclusions: 
\begin{enumerate}
    
    \item The two-meson cloud yields an attractive self-energy for
    all members of the antidecuplet. The observation of attraction is
    consistent with the previous attempts to describe 
        the $\Theta^+$ as a $K\pi N$ state~\cite{Bicudo:2003rw,Kishimoto:2003xy,
    Bicudo:2004cm,Llanes-Estrada:2003us}.
    
    \item It also contributes to the splitting between antidecuplet 
    members, which is only moderately cutoff dependent and provides
    about $20 \%$ of the total splitting to a stronger effect for 
    reasonable values of the cutoff.
    The role played by the two-meson cloud is therefore of
    relevance for a precise understanding of the nature
    of the $\Theta^+$  and the antidecuplet. 
    
    \item The magnitude of $20 \%$ is also
    in agreement quantitatively with 
    the strength of attraction found in the previous 
    study of $BMM$ three-body system~\cite{Llanes-Estrada:2003us}.
    The values of the mass splitting are such that they still leave
    some room for quark correlation effects after 
    the GMO mass splitting coming from
    the mass difference between
    $u$, $d$ and $s$ constituent quarks
    is considered. 
        The contribution to the splitting from the meson cloud is of the 
        same order of magnitude as the one provided by 
        these quark correlations.  
    
    \item From the experimental point of view, it is clear that the
    investigation  of the decay channels into two mesons and a baryon 
    of the resonances $N(1710)$, $\Sigma(1770)$, and $\Xi(1860)$ 
    deserves renewed interest.
\end{enumerate}

\begin{acknowledgments}
    This work is supported by the Japan-Europe (Spain) Research
    Cooperation Program of Japan Society for the Promotion of Science
    (JSPS) and Spanish Council for Scientific Research (CSIC), which
    enabled
    E.O. and M.J.V.V. visit RCNP, Osaka.
    This work is partly supported by DGICYT contract numbers BFM2003-00856,
    FPA 2000-0956, BFM 2002-01003
    and the E.U. EURIDICE network contract no. HPRN-CT-2002-00311.
\end{acknowledgments}

\appendix

\section{Flavor coefficients for $PBMM$ vertices}\label{sec:Coef}

This appendix contains
the flavor coefficients for the tree-level vertices in
the three Lagrangians considered.
The coefficients for $\Theta_{\overline{10}}$, $N_{\overline{10}}$,
$\Sigma_{\overline{10}}$, and $\Xi_{\overline{10}}$ are shown in 
Tables~\ref{tbl:Coef1}, \ref{tbl:Coef2}, ~\ref{tbl:Coef3},
and \ref{tbl:Coef4}, respectively.

\begin{table}[btp]
\centering
\caption{The $C^{(j)}_{B,m_1,m_2}$ flavor coefficients for
the vertex with $\Theta^+_{\overline{10}}$, octet baryons, and
two octet mesons.}
\begin{ruledtabular}
\begin{tabular}{llcccc}
    $P$ & $BMM$
    & $8s$ & $8a$ & $27$  & $M$ \\
    \hline
    $\Theta^+_{\overline{10}}$ & $pK^+\pi^-$ & $-\sqrt{6}$ & $-\sqrt{6}$
      & $\frac{4\sqrt{6}}{5}$ & $-\sqrt{6}\Kpi$ \\
    & $pK^0\pi^0$ & $\sqrt{3}$ & $\sqrt{3}$ 
      & $-\frac{4\sqrt{3}}{5}$ & $\sqrt{3}\Kpi$\\
    & $pK^0\eta$  & $1$        & $-3$
      & $\frac{36}{5}$& $\Kcpi$ \\
    & $nK^+\pi^0$ & $\sqrt{3}$ & $\sqrt{3}$
      & $-\frac{4\sqrt{3}}{5}$ & $\sqrt{3}\Kpi$\\
    & $nK^+\eta$  & $-1$       & $3$
      & $-\frac{36}{5}$ &  $-\Kcpi$ \\
    & $nK^0\pi^+$ & $\sqrt{6}$ & $\sqrt{6}$
      & $-\frac{4\sqrt{6}}{5}$ & $\sqrt{6}\Kpi$\\
    & $\Sigma^+K^0K^0$ & - & - & $-4\sqrt{6}$ & - \\
    & $\Sigma^0K^0K^+$ & - & - & $-8\sqrt{3}$ & - \\
    & $\Sigma^-K^+K^+$ & - & - & $4\sqrt{6}$ & - \\
\end{tabular}
\end{ruledtabular}
\label{tbl:Coef1}
\end{table}

\begin{table}[btp]
\centering
\caption{The $C^{(j)}_{B,m_1,m_2}$ flavor coefficients for
the vertex with $N^+_{\overline{10}}$, octet baryons, and
two octet mesons. Coefficients for $N^0_{\overline{10}}$
are obtained by using isospin symmetry.}
\begin{ruledtabular}
\begin{tabular}{llcccc}
    $P$ & $BMM$ 
    & $8s$ & $8a$ & $27$  & $M$ \\ 
    \hline
    $N^{+}_{\overline{10}}$ & $pK^+K^-$ &$-\sqrt{2}$  & $-\sqrt{2}$ 
      & $\frac{4\sqrt{2}}{5}$ & $-\sqrt{2}\mk$ \\
    & $pK^0\bar{K}^0$ & - & $-2\sqrt{2}$ 
      & $4\sqrt{2}$ & - \\
    & $p\pi^0 \pi^0$ &$\frac{1}{\sqrt{2}}$ & -     
      & $-\frac{2\sqrt{2}}{5}$ & $\frac{1}{\sqrt{2}}\mpi$ \\
    & $p\pi^+\pi^-$ &$\sqrt{2}$ & $\sqrt{2}$ 
      & $-\frac{4\sqrt{2}}{5}$ & $\sqrt{2}\mpi$ \\
    & $p\eta\eta$ &$-\frac{1}{\sqrt{2}}$ & -      
      & $-\frac{18\sqrt{2}}{5}$ &$-\frac{1}{\sqrt{2}}\Kopi$ \\
    & $p\eta\pi^0$ & $-\sqrt{\frac{2}{3}}$ & -
      & $\frac{8\sqrt{6}}{5}$ & $-\sqrt{\frac{2}{3}}\mpi$  \\
    & $n\bar{K}^0K^+$ &$-\sqrt{2}$ & $-\sqrt{2}$
      & $-\frac{16\sqrt{2}}{5}$ & $-\sqrt{2}\mk$ \\
    & $n\pi^+\eta$ &$-\frac{2}{\sqrt{3}}$ & -
      & $\frac{16\sqrt{3}}{5}$ & $-\frac{2}{\sqrt{3}}\mpi$ \\
    & $n\pi^+\pi^0$ & - & $-2$       
      & - & - \\
    & $\Lambda K^+\pi^0$ & $-\sqrt{\frac{3}{2}}$ & $-\sqrt{\frac{3}{2}}$
      & $\frac{2\sqrt{6}}{5}$ & $-\sqrt{\frac{3}{2}}\Kpi$\\
    & $\Lambda K^+\eta$ & $\frac{1}{\sqrt{2}}$ & $-\frac{3}{\sqrt{2}}$
      & $\frac{18\sqrt{2}}{5}$ & $\frac{1}{\sqrt{2}}\Kcpi$\\
    & $\Lambda K^0\pi^+$ & $-\sqrt{3}$ & $-\sqrt{3}$
      & $\frac{4\sqrt{3}}{5}$ & $-\sqrt{3}\Kpi$\\
    & $\Sigma^+K^+\pi^-$ & $\sqrt{2}$ &$\sqrt{2}$
      & $-\frac{4\sqrt{2}}{5}$ & $\sqrt{2}\Kpi$ \\
    & $\Sigma^+K^0\eta$ &$-\frac{1}{\sqrt{3}}$ & $\sqrt{3}$ 
      & $\frac{28\sqrt{3}}{5}$ & $-\frac{1}{\sqrt{3}}\Kcpi$\\
    & $\Sigma^+K^0\pi^0$ & $-1$ & $-1$
      & $-\frac{36}{5}$ &$-\Kpi$  \\
    & $\Sigma^0K^+\pi^0$ &$\frac{1}{\sqrt{2}}$ & $\frac{1}{\sqrt{2}}$
      & $-\frac{22\sqrt{2}}{5}$ & $\frac{1}{\sqrt{2}}\Kpi$\\
    & $\Sigma^0K^0\pi^+$ & $1$ & $1$
      & $\frac{36}{5}$ &  $\Kpi$\\
    & $\Sigma^0K^+\eta$ &$-\frac{1}{\sqrt{6}}$ & $\sqrt{\frac{3}{2}}$
      & $\frac{14\sqrt{6}}{5}$ & $-\frac{1}{\sqrt{6}}\Kcpi$ \\
    & $\Sigma^-K^+\pi^+$ & - & -
      & $-8\sqrt{2}$ & - \\
    & $\Xi^-K^+K^+$ & - & -
      & $4\sqrt{2}$ & - \\
    & $\Xi^0K^+K^0$ & - & -
      & $4\sqrt{2}$ & - \\
\end{tabular}
\end{ruledtabular}
\label{tbl:Coef2} 
\end{table}

\begin{table}[btp]
\centering
\caption{The $C^{(j)}_{B,m_1,m_2}$ flavor coefficients for
the vertex with $\Sigma^+_{\overline{10}}$, octet baryons, and
two octet mesons. Coefficients for $\Sigma^0_{\overline{10}}$
and $\Sigma^-_{\overline{10}}$
are obtained by using isospin symmetry.}
\begin{ruledtabular}
\begin{tabular}{llcccc}
    $P$ & $BMM$ 
    & $8s$ & $8a$ & $27$ & $M$ \\ 
    \hline
    $\Sigma^+_{\overline{10}}$ & $p\pi^+K^-$ & $-\sqrt{2}$
      & $-\sqrt{2}$ & $\frac{4\sqrt{2}}{5}$ & $-\sqrt{2}\Kpi$ \\
    & $p\pi^0\bar{K}^0$ & $1$  & $1$  
      & $-\frac{24}{5}$ & $\Kpi$ \\
    & $p\bar{K}^0\eta$ & $\frac{1}{\sqrt{3}}$ & $\sqrt{3}$  
      & $\frac{32\sqrt{3}}{5}$ & $\frac{1}{\sqrt{3}}\Kcpi$ \\
    & $n\bar{K}^0\pi^+$ & - & -
      & $-4\sqrt{2}$ & - \\
    & $\Lambda\bar{K}^0K^+$ & $-\sqrt{3}$ & $-\sqrt{3}$ 
      & $-\frac{16\sqrt{3}}{5}$ & $-\sqrt{3}\mk$ \\
    & $\Lambda\pi^+\eta$ & $-\sqrt{2}$ & -
      & $\frac{24\sqrt{2}}{5}$ &    $-\sqrt{2}\mpi$ \\
    & $\Lambda\pi^+\pi^0$  & - & $-\sqrt{6}$ 
      & - & - \\
    & $\Sigma^+K^+K^-$ &$-\sqrt{2}$ & $-\sqrt{2}$ 
      & $\frac{4\sqrt{2}}{5}$ & $-\sqrt{2}\mk$ \\
    & $\Sigma^+\pi^+\pi^-$ &$\sqrt{2}$ & $\sqrt{2}$ 
      & $-\frac{4\sqrt{2}}{5}$ & $\sqrt{2}\mpi$ \\
    & $\Sigma^+\eta\eta$ & $-\frac{1}{\sqrt{2}}$ & -      
      & $\frac{12\sqrt{2}}{5}$ &  $-\frac{1}{\sqrt{2}}\Kopi$ \\
    & $\Sigma^+\eta\pi^0$ & $-\sqrt{\frac{2}{3}}$ & -   
      & $-\frac{12\sqrt{6}}{5}$ & $-\sqrt{\frac{2}{3}}\mpi$ \\
    & $\Sigma^+\pi^0\pi^0$ & $\frac{1}{\sqrt{2}}$ & -
      & $\frac{8\sqrt{2}}{5}$ & $\frac{1}{\sqrt{2}}\mpi$  \\
    & $\Sigma^+K^0\bar{K}^0$ & - &$-2\sqrt{2}$
      & $-4\sqrt{2}$ & - \\
    & $\Sigma^0\bar{K}^0K^+$ & $1$ & $1$ 
      & $-\frac{24}{5}$ & $\mk$ \\
    & $\Sigma^0\eta\pi^+$ & $\sqrt{\frac{2}{3}}$ & -     
      & $\frac{12\sqrt{6}}{5}$ &  $\sqrt{\frac{2}{3}}\mpi$ \\
    & $\Sigma^0\pi^+\pi^0$ & - & $\sqrt{2}$ 
      & $-4\sqrt{2}$ & - \\
    &$\Sigma^-\pi^+\pi^+$ & - & -
      & $-4\sqrt{2}$ & - \\
    & $\Xi^0K^+\pi^0$ & $1$ & $1$
      & $-\frac{24}{5}$ & $\Kpi$ \\
    & $\Xi^0K^0\pi^+$ & $\sqrt{2}$ &$\sqrt{2}$ 
      & $\frac{16\sqrt{2}}{5}$ & $\sqrt{2}\Kpi$\\
    & $\Xi^0K^+\eta$ & $-\frac{1}{\sqrt{3}}$ & $-\sqrt{3}$ 
      & $\frac{8\sqrt{3}}{5}$ & $-\frac{1}{\sqrt{3}}\Kcpi$ \\
    & $\Xi^-K^+\pi^+$ & - & - 
      & $8\sqrt{2}$ & - \\
\end{tabular}
\end{ruledtabular}
\label{tbl:Coef3} 
\end{table}

\begin{table}[btp]
\centering
\caption{The $C^{(j)}_{B,m_1,m_2}$ flavor coefficients for
the vertex with $\Xi^+_{\overline{10}}$, octet baryons, and
two octet mesons. Coefficients for $\Xi^0_{\overline{10}}$,
$\Xi^-_{\overline{10}}$ and $\Xi^{--}_{\overline{10}}$
are obtained by using isospin symmetry.}
\begin{ruledtabular}
\begin{tabular}{llcccc}
    $P$ & $BMM$ 
    & $8s$ & $8a$ & $27$ & $M$ \\ 
    \hline
    $\Xi^+_{\overline{10}}$ & $\Sigma^+\pi^+K^-$ & $-\sqrt{6}$ & $-\sqrt{6}$ 
      & $\frac{4\sqrt{6}}{5}$ & $-\sqrt{6}\Kpi$  \\
    & $\Sigma^+\pi^0\bar{K}^0$ & $\sqrt{3}$ & $\sqrt{3}$
      & $\frac{16\sqrt{3}}{5}$ & $\sqrt{3}\Kpi$ \\
    & $\Sigma^+\bar{K}^0\eta$ & $1$ & $3$        
      & $-\frac{24}{5}$ & $\Kcpi$ \\
    & $\Sigma^0\bar{K}^0\pi^+$ & - & -       
      & $-4\sqrt{3}$ & - \\
    & $\Lambda\bar{K}^0\pi^+$ & - & -       
      & $-12$ & - \\
    & $p\bar{K}^0\bar{K}^0$ & - & -       
      & $-4\sqrt{6}$ & - \\
    & $\Xi^0\bar{K}^0 K^+$ & $\sqrt{6}$ & $\sqrt{6}$
      & $-\frac{4\sqrt{6}}{5}$ & $\sqrt{6}\mk$ \\
    & $\Xi^0\eta\pi^+$ & $2$ & - 
      & $\frac{12}{5}$ &  $2\mpi$ \\
    & $\Xi^0\pi^+\pi^0$ & - & $2\sqrt{3}$ 
      & $-4\sqrt{3}$ & -\\
    & $\Xi^-\pi^+\pi^+$ & - & -
      & $4\sqrt{6}$& -\\
\end{tabular}
\end{ruledtabular}
\label{tbl:Coef4} 
\end{table}

\section{Self-energy Formulae}\label{sec:Formulae}

Below are the formulae for calculating the self-energies
as described in
Eq.~\eqref{eq:selfenergy}. In the isospin symmetric limit,
\begin{equation}
    \Sigma^{(j)}_{P}
    = \left(F^{(j)}\right)^2
    \sum_{\alpha}
    I^{(j)} (\alpha) D^{(j)}_{P,\alpha} ,
    \label{eq:selfenergy2}
\end{equation}
with $\alpha$ being the $BMM$ channel in the isospin basis, such as
$NK\pi$, $NK\eta$, etc., and $D^{(j)}$ are expressed
as the sum of the $(C^{(j)})^2$.
In Table IX--XII, we show the $D^{(j)}$ coefficients.
For the $27$ and $M$ cases,
following the procedure in Sec.~\ref{subsec:27M},
we set $g^{8s}=g^{27}=g^{M}=1.88$ and take
\begin{equation}
    a\mathcal{L}^{8s}+b\mathcal{L}^{27} ,
    \quad
    b=-\frac{5}{4}(1-a)
    \label{eq:Lagcomb1_a}
\end{equation}
and
\begin{equation}
    a \mathcal{L}^{8s} + b\mathcal{L}^{M} , \quad
    b = \frac{f^2}{m_{\pi}^2}(1-a) ,
    \label{eq:Lagcomb2_a}
\end{equation}
In these cases, $D^{j}$ are defined as
\begin{equation}
    \Sigma^{(j)}_{P}
    = \left(F^{(8s)}\right)^2
    \sum_{\alpha}
    I^{(j)} (\alpha) D^{(j)}_{P,\alpha} ,
    \label{eq:selfenergy3}
\end{equation}
for $(j) = 8s+27, 8s+M$.
One can easily check that when $a=1$, $b=0$,
$D^{(j)}_{P,\alpha}$ for $(j)=8s+27, 8s+M $
becomes $D^{(8s)}_{P,\alpha}$.

\begin{table}[btp]
\centering
\caption{The $D^{(j)}_{\alpha}$ coefficients for
the $\Theta_{\overline{10}}$ self-energies. }
\begin{ruledtabular}
\begin{tabular}{llllll}
    $P$ & $\alpha$ 
    & $8s$ & $8a$ & $8s + 27$ & $8s + M$ \\ 
    \hline
    $\Theta_{\overline{10}}$ & $NK\pi$ & 18 & 18 & 18
      & $18(a+\Kpi b)^2$ \\
    & $NK\eta$ & 2 & 18 & $2(a+\frac{36}{5}b)^2$
    & $2(a+\Kcpi b)^2$  \\
    & $\Sigma K\pi$ & - & - & $576b^2$ & -\\
\end{tabular}
\end{ruledtabular}
\label{tbl:Dcoef1} 
\end{table}

\begin{table}[btp]
\centering
\caption{The $D^{(j)}_{\alpha}$ coefficients for
the $N_{\overline{10}}$ self-energies. }
\begin{ruledtabular}
\begin{tabular}{llllll}
    $P$ & $\alpha$ 
    & $8s$ & $8a$ & $8s + 27$ & $8s + M$ \\ 
    \hline
    $N_{\overline{10}}$ & $NK\bar{K}$ & 4 & 12       
      & $2+2(a+\frac{16}{5}b)^2-32b^2$ 
      & $4(a+\mk b)^2$\\
    & $N\pi\pi$ & 3 & 6 & 3 & $3(a+\mpi b)^2$\\
    & $N\pi\eta$ & 2 & - & $2(-a+\frac{24}{5}b)^2$ 
      & $2(a+\mpi b)^2$\\
    & $N\eta\eta$ & 1 & - & $(a+\frac{36}{5})^2$ 
      & $(a+\Kopi b)^2$\\
    & $\Lambda K\pi$ & $\frac{9}{2}$ & $\frac{9}{2}$ 
      & $\frac{9}{2}$ & $\frac{9}{2}(a+\Kpi b)^2$ \\
    & $\Lambda K\eta$ & $\frac{1}{2}$ & $\frac{9}{2}$
      & $\frac{1}{2}(a+\frac{36}{5})^2$
      & $\frac{1}{2}(a+\Kcpi b)^2$ \\
    & $\Sigma K\pi$ & $\frac{9}{2}$ & $\frac{9}{2}$
      & $2+2(a+\frac{36}{5}b)^2$ & $(a+\Kpi b)^2$ \\
    &&&& $+\frac{1}{2}(a-\frac{44}{5})^2+128b^2$\\
    & $\Sigma K\eta$ & $\frac{1}{2}$ & $\frac{9}{2}$
      & $\frac{1}{2}(-a+\frac{84}{5})^2$
      & $\frac{1}{2}(a+\Kcpi b)^2$ \\
    & $\Xi KK$ & - & - & $96b^2$ & -\\ 
\end{tabular}
\end{ruledtabular}
\label{tbl:Dcoef2} 
\end{table}

\begin{table}[btp]
\centering
\caption{The $D^{(j)}_{\alpha}$ coefficients for
the $\Sigma_{\overline{10}}$ self-energies. }
\begin{ruledtabular}
\begin{tabular}{llllll}
    $P$ & $\alpha$ 
    & $8s$ & $8a$ & $8s + 27$ & $8s + M$ \\ 
    \hline
    $\Sigma_{\overline{10}}$ & $N\bar{K}\pi$ & 3 & 3 
      & $2+(a-\frac{24}{5}b)^2+32b^2$ 
      & $3(a+\Kpi b)^2$\\
    & $N\bar{K}\eta$ & $\frac{1}{3}$ & 3 
      & $\frac{1}{3}(a+\frac{96}{5}b)^2$ 
      & $\frac{1}{3}(a+\Kcpi)^2$ \\
    & $\Lambda K\bar{K}$ & 3 & 3 
      & $3(a+\frac{16}{5}b)^2$
      & $3(a+\mk b)^2$ \\
    & $\Lambda \pi\eta$ & 2 & - 
      & $2(a-\frac{24}{5}b)^2$
      & $2(a+\mpi b)^2$\\
    & $\Lambda \pi\pi$ & - & 6 & - & - \\
    & $\Sigma K\bar{K}$ & 3 & 11 
      & $2+(a-\frac{24}{5}b)^2+32b^2$
      & $3(a+\mk b)^2$ \\
    & $\Sigma \pi\pi$ & 3 & 4 
      & $2+(a+\frac{16}{5})^2+96b^2$
      & $(a+\mpi b)^2$\\
    & $\Sigma \pi\eta$ & $\frac{4}{3}$ & -
      & $\frac{4}{3}(a+\frac{36}{5})^2$
      & $\frac{4}{3}(a+\mpi b)^2$\\
    & $\Sigma \eta\eta$ & 1 & - 
      & $(a-\frac{24}{5}b)^2$ 
      & $(a+\Kopi b)^2$\\
    & $\Xi K\pi$ & 3 & 3 
      & $2(a+\frac{16}{5})^2$ 
      & $3(a+\Kpi b)^2$ \\
    &&&& $+(a-\frac{24}{5})^2+128b^2$ \\
    & $\Xi K\eta$ & $\frac{1}{3}$ & 3 
      & $\frac{1}{3}(a-\frac{24}{5}b)^2$
      & $\frac{1}{3}(a+\Kcpi b)^2$ \\
\end{tabular}
\end{ruledtabular}
\label{tbl:Dcoef3} 
\end{table}

\begin{table}[btp]
\centering
\caption{The $D^{(j)}_{\alpha}$ coefficients for
the $\Xi_{\overline{10}}$ self-energies. }
\begin{ruledtabular}
\begin{tabular}{llllll}
    $P$ & $\alpha$ 
    & $8s$ & $8a$ & $8s + 27$& $8s + M$  \\ 
    \hline
    $\Xi_{\overline{10}}$ & $\Sigma \bar{K}\pi$ & 9 & 9
      & $6+3(a+\frac{16}{5}b)^2+48b^2$  
      & $9(a+\Kpi b)^2$ \\
    & $\Sigma \bar{K}\eta$ & 1 & 9 
      & $(a-\frac{24}{3}b)^2$ 
      & $(a+\Kcpi b)^2$ \\ 
    & $\Xi K\bar{K}$ & 6 & 6 & 6
      & $6(a+\mk b)^2$ \\ 
    & $\Xi \pi\eta$ & 4 & - 
      & $4(a+\frac{6}{5}b)^2$ 
      & $4(a+\mpi b)^2$\\
    & $\Xi \pi\pi$ & - & 12 & $240b^2$ &-\\
    & $\Lambda \bar{K}\pi$ & - & - & $144b^2$ &- \\ 
    & $N \bar{K}\bar{K}$ & - & - & $192b^2$ &- \\ 
\end{tabular}
\end{ruledtabular}
\label{tbl:Dcoef4} 
\end{table}

\end{document}